\documentclass[12pt]{aastex62} 
\usepackage{amsmath}
\usepackage{footmisc}
\DefineFNsymbols{mySymbols}{{\ensuremath\dagger}}
\setfnsymbol{mySymbols}

\begin{document} 
 
\received{} 
\accepted{} 
 
\title{Modeling Stellar Oscillations and Granulation in Radial Velocity Time Series: A Fourier-based Method}  

\author{Zhao Guo}
\altaffiliation{Current: DAMTP, University of Cambridge, zg281@cam.ac.uk}
\affil{Department of Astronomy, The Pennsylvania State University, 525 Davey Lab, University Park, PA 16802, USA} 
\affil{Center for Exoplanets and Habitable Worlds, 525 Davey Lab, The Pennsylvania State University, University Park, PA, 16802, USA}

\author{Eric B. Ford}
\affil{Department of Astronomy, The Pennsylvania State University, 525 Davey Lab, University Park, PA 16802, USA} 
\affil{Center for Exoplanets and Habitable Worlds, 525 Davey Lab, The Pennsylvania State University, University Park, PA, 16802, USA}
\affil{Center for Astrostatistics, 525 Davey Laboratory, The Pennsylvania State University, University Park, PA, 16802, USA}
\affil{Institute for Computational \& Data Sciences, The Pennsylvania State University, University Park, PA, 16802, USA}

\author{Dennis Stello}
\affil{School of Physics, The University of New South Wales, Sydney NSW 2052, Australia}

\author{Jacob K. Luhn}
\affil{Department of Astronomy, The Pennsylvania State University, 525 Davey Lab, University Park, PA 16802, USA} 
\affil{Center for Exoplanets and Habitable Worlds, 525 Davey Lab, The Pennsylvania State University, University Park, PA, 16802, USA}

\author{Suvrath Mahadevan}
\affil{Department of Astronomy, The Pennsylvania State University, 525 Davey Lab, University Park, PA 16802, USA} 
\affil{Center for Exoplanets and Habitable Worlds, 525 Davey Lab, The Pennsylvania State University, University Park, PA, 16802, USA}

\author{Arvind F. Gupta}
\affil{Department of Astronomy, The Pennsylvania State University, 525 Davey Lab, University Park, PA 16802, USA} 
\affil{Center for Exoplanets and Habitable Worlds, 525 Davey Lab, The Pennsylvania State University, University Park, PA, 16802, USA}

\author{Jie Yu}
\affil{Max Planck Institute for Solar System Research, Justus-von-Liebig-Weg 3, 37077 Gottingen, Germany} 

%\slugcomment{09/06/2020} 
%\paperid{}

%%%%%%%%%%%%%%%%%%%%%%%%%%%%%%%%%%%%%%%%%%%%%%%%%%%%%%%%%%%%%% 

\begin{abstract} 

Tens of thousands of solar-like oscillating stars have been observed by space missions. Their photometric variability in the Fourier domain can be parameterized by a sum of two super-Lorentizian functions for granulation and a Gaussian-shaped power excess for oscillation. The photometric granulation/oscillation parameters scale with stellar parameters and they can also make predictions for corresponding parameters in radial velocity measurements. Based on scaling relations, we simulate realistic radial velocity time series and examine how the root-mean-square scatter of radial velocity measurements varies with stellar parameters and different observation strategies such as the length of integration time and gaps in the time series.
Using stars with extensive spectroscopic observations from the spectrographs (SONG and HARPS), we measure the granulation amplitude and timescale from the power spectrum of the radial velocity time series.
We compare these measurements with literature values based on {\it Kepler} photometry. We find that the granulation amplitude in radial velocity can be well predicted from the photometry and scaling relations. Both granulation timescales in radial velocity agree with those predicted from photometry for giants and sub-giants. However, for main-sequence stars, only one granulation timescale in radial velocity is in agreement with the photometric-based values, while the other timescale generally lies at lower frequencies compared to the result of photometry. In conclusion, we show the photometric scaling relations from {\it Kepler} photometry and the scaling relationship to Doppler observations can be very useful for predicting the photometric and radial velocity stellar variabilities due to stellar granulation and oscillation.

\end{abstract}

%\keywords{stars:binaries: 
%}
 
%%%%%%%%%%%%%%%%%%%%%%%%%%%%%%%%%%%%%%%%%%%%%%%%%%%%%%%%%%%%%%% 

\section{Introduction}

Stellar variability of solar-type stars is driven by a variety of processes including granulation, oscillation, rotational modulation, and stellar magnetic activities, with typical timescales from minutes to years.
Thanks to space missions such as {\it Kepler}, significant advances have been made in our understanding of stellar granulation and oscillation. This is primarily from studying the Fourier (power) spectrum of photometric variability. The amplitude and timescale of the granulation has been found to correlate with the frequency of maximum acoustic oscillation power $\nu_{max}$ (Kallinger et al.\ 2014, hereafter K14). Typically, two components of granulation are found. Other methods to infer the stellar parameters (e.g., surface gravity) also use the power spectrum, such as the autocorrelation function (ACF)-timescale method in Kallinger et al.\ (2016) and the frequency-filtered power spectrum method in Bugnet et al. (2018). Blancato et al.\ (2020) used light curves, their ACF, and their power spectra as input to neural networks and tried to infer stellar parameters including rotational periods, for a wide range of stars in the HR diagram. 

The effect of stellar variability on radial velocity (RV) Doppler observations is complex, especially when magnetic activities are considered. Apart from the instrumental effect, the observed velocity of a star is the sum of the star's center of mass velocity, measurement noise (typically assumed to be drawn from a known distribution with zero mean and known variance), and intrinsic stellar variability (signal to asteroseismologists but noise to planet hunters). The commonly used concept of `RV jitter' (e.g., Yu et al.\ 2018a; Tayar et al.\ 2019) further assumes that the stellar variability term is independent and identically distributed (i.i.d.), so the jitter effectively contributes an excess noise term with no correlation between separate RV observations. The RV Fourier spectrum of stellar variability has not been given commensurable attention compared to the photometric Fourier spectrum (see Dumusque et al. 2011, 2016; Delisle et al.\ 2020 for exceptions). Here we use a simple, empirical, observation-driven approach to predict the Fourier spectrum of RV observations due to short-term stellar variability from granulation and oscillations. Our model replaces the i.i.d. RV jitter model with a power spectrum model for these intrinsic stellar variability terms. The power density spectrum (PDS), which is the squared Fourier amplitude multiplied by the time span of the time series, visualizes how both granulation and oscillation contribute to the `stellar jitter', which impedes the detection of exoplanet and which we aim to mitigate. Although oscillations of solar-like stars are in a narrow frequency range (the power excess around $\nu_{max}$), the frequency dependence of granulation spans several orders of magnitudes from minutes to days. The commonly referred RV jitter due to stellar variability is actually the integral of our PDS model from the minimum frequency to the Nyquist frequency. Our PDS model for stellar variability accounts for a frequency-dependence of granulation and oscillation,  as well as the correlation between the RV measurements within a night. Furthermore, it can also be scaled to different stars using their effective temperature and surface gravity.

{\it Kepler} observations of a large sample of solar-like oscillating stars, ranging from main sequence stars to red giants, have revealed empirical scaling relations for the oscillation and granulation background signatures in the power spectrum. In the Fourier domain, stellar oscillations manifests itself as a roughly Gaussian-shaped power excess centered at the frequency of maximum acoustic power ($\nu_{max}$), with a characteristic width of ($\sigma_{env}$) and height ($H_{osc}$). Working in the Fourier domain, Chaplin et al.\ (2019) explored the effect from acoustic oscillations of different integration time on the RMS scatter of RV time series. They found that, after high-pass filtering the RV time series, the RMS scatter is a decreasing function of the integration time $\Delta t_c$, with dips corresponding to the integer times of the oscillation period ($1/\nu_{max}$). Their work presents the method to reduce the effect of stellar oscillations of planet surveys. However, they did not consider the effect of granulation.

In Section 2, we describe a parameterized model that can be used to describe the granulation and oscillation in the Fourier spectrum. In Section 3, we use that model to perform the inverse Fourier transform and simulate the RV time series for stars with different stellar parameters. In Section 4, using these simulated RV data, we explore different observation strategies to mitigate RV jitter induced by stellar granulation and oscillations. We discuss the caveats and future prospects in Section 5.

In the Appendix we perform the background fitting to the power spectrum of RV time series. This is commonly done with photometry, e.g., based on {\it Kepler} light curves (K14), but relatively rare in Doppler observations (Kjeldsen \& Bedding 2011, Dumusque et al.\ 2011). We describe the functional form used in the fitting and the method for the model comparison. We also present the measured granulation amplitude and timescale and compare them with theoretical predictions from the scaling relations.

\section{Stellar granulation and oscillations in the Fourier space}    

In the exoplanet community, some form of a periodogram which shows the significance of a periodic signal as a function of the putative orbital period of the RVs, is commonly used to find the orbital period of planets inducing a reflex motion of the star.

Apart from stellar rotation and magnetic activity, stellar variability mainly arises from granulation and oscillations, which are both related to the surface convection. In asteroseismology, stellar variability is routinely examined in Fourier space. The granulation signature is typically parameterized by two super-Lorentzian functions (the first term in the parenthesis of eq.\ (1)), spanning a broad range of frequencies (see green and blue dashed lines in Figure 1 for illustration). The effects of stellar oscillations on the PDS for a series of either photometric or RV measurements of a star can be well approximated by a Gaussian-shaped bump (the second term in the parenthesis of eq.\ 1) centered at $\nu_{max}$, with a characteristic width ($\sigma_{env}$). 
Other than the above, there is also a frequency-independent term (white noise $W$ in eq.\ 1), which can be seen as a flat base at the very high-frequency part of the PDS (see the brown flat lines in Figure 1). Rotational modulation can appear as a dominant frequency peak at the rotation period, as well as several lower-amplitude peaks at orbital harmonic frequencies. As the  amplitude of magnetic activity in the form of rotationally modulated variability is more difficult to model, we do not consider these effects here (see However, Meunier et al.\ 2019).

\subsection{Photometric variability in the Fourier space}   

Following K14, we model the photometric PDS of solar-like oscillating stars by eq. (1). There are strong empirical relationships for $a$ (granulation amplitude) and $b$ (granulation characteristic frequency) as a function of $\nu_{max}$ (see eq.\ 3, 4, 5 below) and similarly, for the oscillation bump parameters $\sigma_{env}$ and $H_{osc}$ as introduced earlier. In addition, a frequency-independent white noise term $W$ is also required in the PDS model.

The mean PDS is modeled by:

\onecolumngrid 

\begin{equation}
\mathcal{P}(\nu;\Theta)=W+ \eta(\nu)^2\left( \sum_{i=1,2}\frac{\xi a_{i}^2/b_i}{1+(\nu/b_i)^4}+H_{osc}exp{\left[-\frac{(\nu-\nu_{max})^2}{2\sigma_{env}^2} \right]}\right),
\end{equation}
with $\Theta=(W, a_1,b_1, a_2, b_2, H_{osc}, \nu_{max}, \sigma_{env})$, $\xi=2\sqrt{2}/\pi$ and $\eta(\nu)=sinc(\frac{\pi}{2}\frac{\nu}{\nu_{Nq}})$. The $\eta(\nu)$ is an attenuation term, taking into account the finite exposure time in realistic time series. For a time series with a regular spacing of $\Delta t$, the corresponding Nyquist frequency is defined as $\nu_{Nq}=1/(2\Delta t)$. It is the highest frequency that can be probed by this regularly-spaced  time series.

\twocolumngrid

Following Kjeldsen \& Bedding (1995) and Basu \& Chaplin (2017, hereafter BC17):
the frequency at maximum power of oscillations follows:
$\nu_{max} \propto gT_{\rm eff}^{-0.5}$. Scaled to the Sun, we have: 
\begin{equation}
\nu_{max} = 3140 \left(\frac{T_{\rm eff}}{5777K}\right)^{-0.5}\left(\frac{g}{g_{\odot}}\right)^{1} (\mu Hz).
\end{equation}

Following K14, the granulation characteristic frequencies (the 'knee' frequency in the Super-Lorentzian) for the two components scale with $\nu_{max}$ as:

\begin{equation}
b_1=0.317\nu_{max}^{0.97} 
\end{equation}
and
\begin{equation}
b_2=0.948\nu_{max}^{0.992}
\end{equation}
The two relations above can be derived by considering that the convection cell covers a vertical distance that is proportional to the pressure scale height $H_p$ at a sound speed $c_s$, so that $b \propto  c_s/H_p \propto g/\sqrt{T_{\rm eff}} \propto \nu_{max}$ (Kjeldsen \& Bedding 2011; K14; BC17).

K14 found that the two granulation components have similar amplitude: $a_1 \approx a_2=a_{gran}$. We thus assume $a_1=a_2=a_{gran}$. 
The granulation amplitudes in {\it Kepler} photometry are related by: 
\begin{equation}
a_{gran} (ppm)=3382\nu_{max}^{-0.609}.
\end{equation}

In addition, the pulsation and granulation amplitudes are related by:

\begin{equation}
a_{gran}(ppm)=4.57 a_{puls}^{0.855},
\end{equation}

where the pulsation amplitude $a_{puls}$ is related to the height $H_{osc}$ and width $\sigma_{env}$ of the Gaussian excess in the PDS by:
\begin{equation}
a_{puls}(ppm)= (\sigma_{env} H_{osc} \sqrt{2\pi})^{0.5} 
\end{equation}.

We can rearrange equation 7 and express $H_{osc}$ explicitly:  
\begin{equation}
H_{osc} (ppm^2/\mu Hz)=\sqrt{\frac{1}{2\pi\sigma_{env}^2}}\left(\frac{a_{gran}}{4.57}\right)^{\frac{2}{0.855}}
\end{equation}
\ 

As for the width of the oscillation power excess, BC17 showed that  
\begin{equation}
\sigma_{env} (\mu Hz)=0.174\nu_{max}^{0.88} \ .
\end{equation}
% and $\sigma_{env}=0.5\nu_{max}$ ($\mu Hz$) for main-sequence or subgiant stars. 

In the end, $b_1, b_2, a_1,a_2, H_{osc}, \sigma_{env}$ can all be expressed as functions of $\nu_{max}$ via Equations 3, 4, 5, 8, 9, and $\nu_{max}$ can be measured empirically from space-based photometry or can be predicted from stellar parameters ($T_{\rm eff}$, $\log g$). Thus, we can generate a parameterized PDS model for stars at different locations of the ($T_{\rm eff}$-$\log g$) diagram. Metallicity also affects the granulation and oscillation amplitude. Yu et al.\ (2018b) (see also Corsaro et al. 2017) showed that metal-rich red giant stars have larger oscillation and granulation power than metal-poor counterparts. For simplicity, we neglect this effect but keep in mind that the amplitudes can vary by about 10-20\% if [Fe/H] changes from -0.7 to 0.5 (Yu et al. 2018b).

\clearpage

\onecolumngrid 

%psdflow.pptx  Desktop/suvrath/
\begin{figure}[htb!]
\begin{center} 
{\includegraphics[height=10cm]{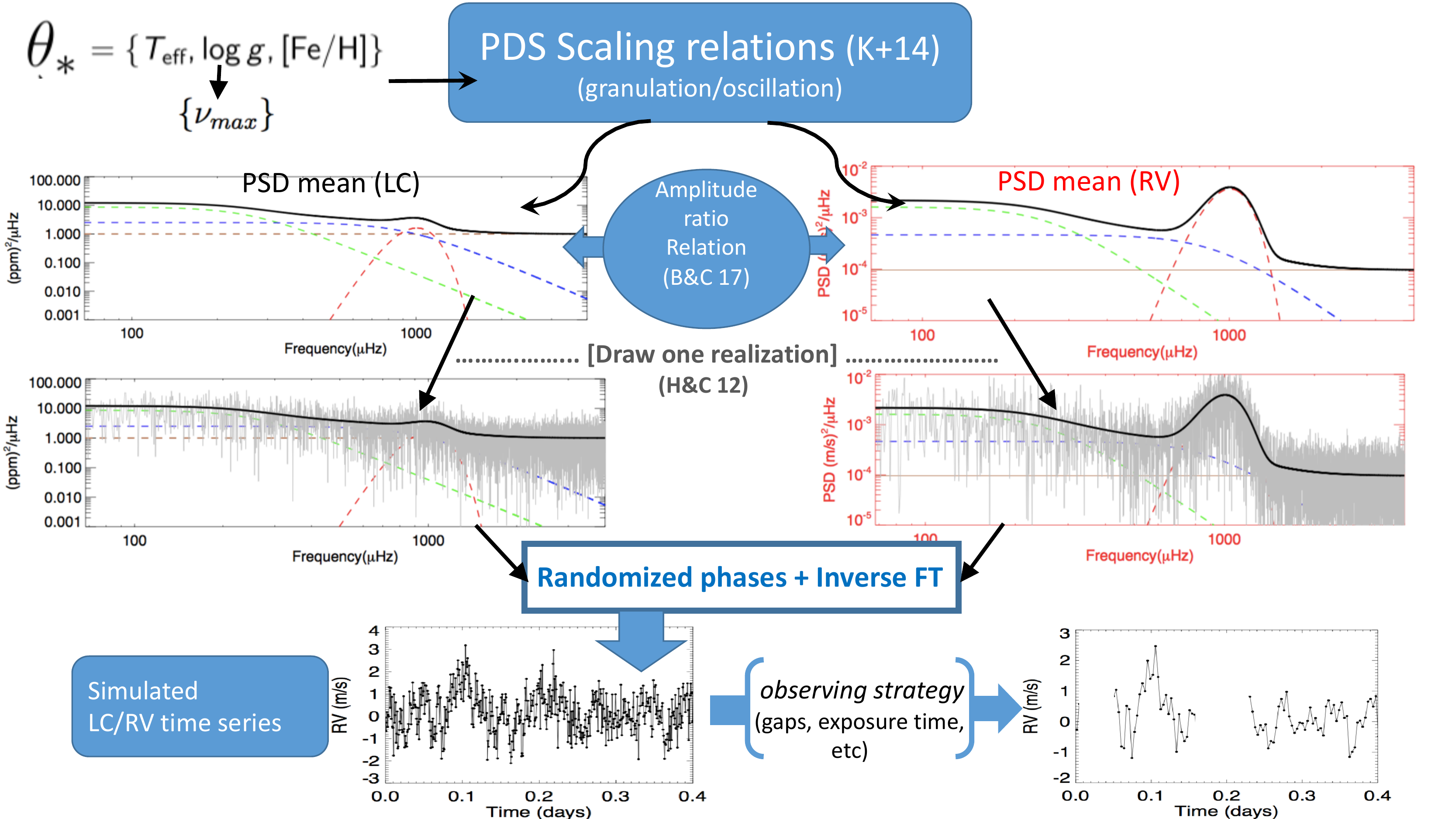}} 
\end{center} 
\caption{The flow chart of simulating photometric and Doppler time series from the stellar power density spectrum(PDS). \textbf{Starting from the upper left and follwing the direction of arrows}, the stellar parameters $\theta_{*}$ is used to generated the mean and observed noisy photometric PDS (labeled as LC for light curves) and Doppler PDS (labeled as RV). Finally, the PDS is inverse-Fourier transformed to generate the simulated RV time series, with observing gaps introduced in the lower right panel.}
\end{figure}

\begin{figure}[h!]
\begin{center} 
{\includegraphics[height=12cm,angle=0]{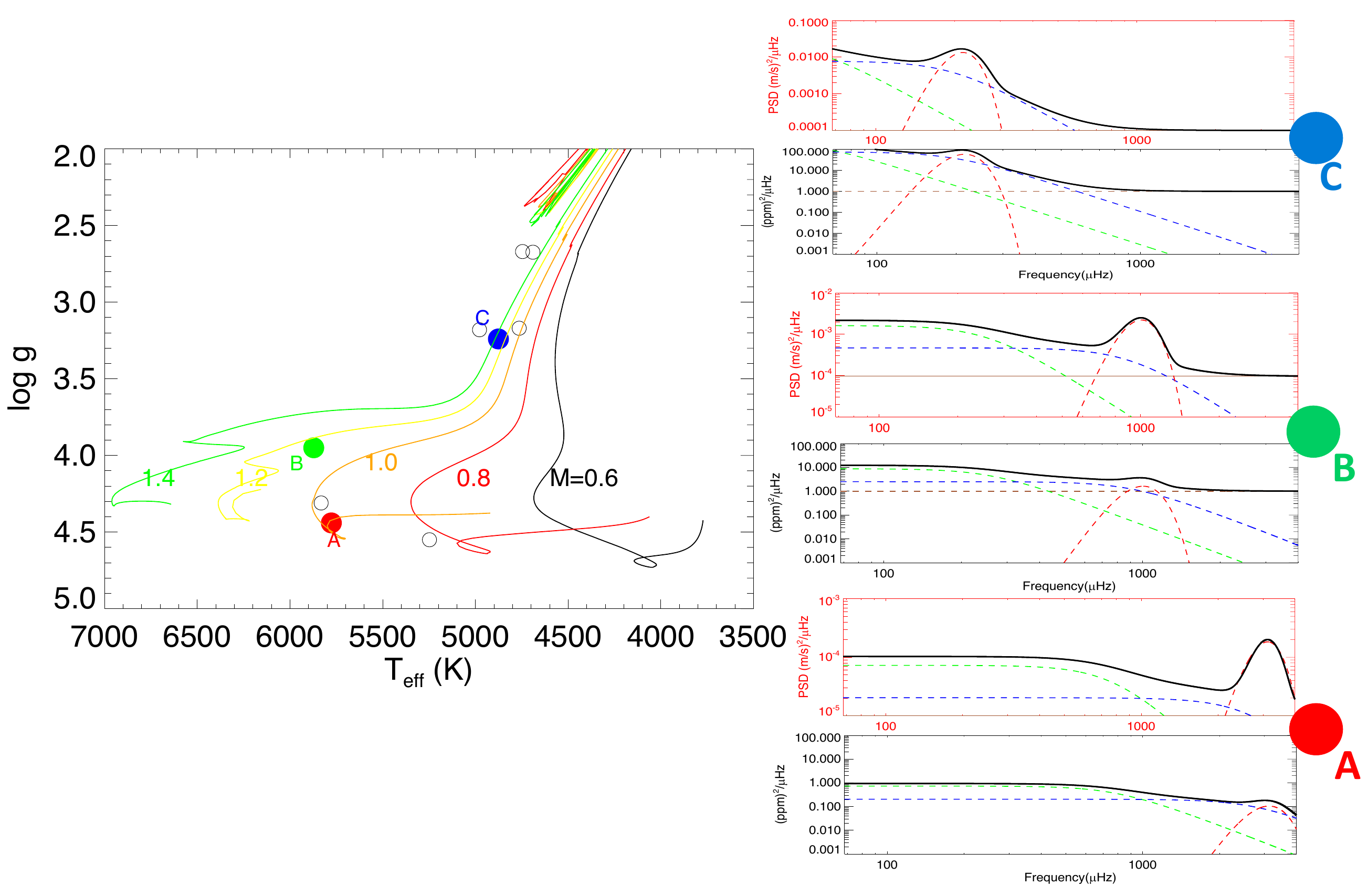}} 
\end{center} 
\caption{The positions of three representative stars on the $\log g-T_{\rm eff}$ diagram (filled circles) and their simualted power density spectrum (Black: Photometric PDS in ${(ppm)}^2/ \mu Hz$; Red: Doppler PDS in ${(m/s)}^2/ \mu Hz$). A: main sequence, the Sun  ($T_{\rm eff}=5777K, \log g=4.44, \nu_{max}=3140\mu Hz$); B: sub-giant, $\beta$ Hyi ($T_{\rm eff}=5872K, \log g=3.95,\nu_{max}=1020 \mu Hz$; C: giant, $\kappa$ CrB ($T_{\rm eff}=4986K, \log g=3.24,\nu_{max}=213 \mu Hz$). Note that a Gaussian-white-noise component is added to the simulated PDS, shown as the brown horizontal lines. It can be seen that the Gaussian-shaped oscillation bump is more dominant in the Doppler PDS. The left panel also shows the positions of additional stars (open circles) studied in the Appendix.}
\end{figure}

\twocolumngrid

\subsection{RV variability in Fourier space} 

We assume that the PDS of the RV time series can be described reasonably by the same parameterized model in eq. (1). Although this is primarily motivated from the {\it Kepler} photometric observations, it has been used for the Doppler observations of the Sun as well (Lefebvre et al.\ 2008). This assumption will be further examined and discussed in the Appendix. 

Based on the photometric scaling relations in Sec. 2.1 for granulation and oscillation, we can convert the photometric variability to the variability in RVs. Following Kjeldsen \& Bedding (1995) and BC17, the granulation amplitude ratio in photometric bolometric observations measured in part per million ($ppm$) and Doppler observations ($m/s$) can be expressed as: 
$a_{gran}(ppm)/a_{gran}(m/s) \propto T_{\rm eff}^{-32/9}g^{2/9}$ This relation is primarily due to the fact that the photometric perturbation is proportional to the temperature perturbation in the stellar photosphere.
Calibrated against the Sun, we have the following amplitude ratio relation:

\begin{equation}
r_{gran,bolo}=\frac{a_{gran,bolo}(ppm)}{a_{gran}(m/s)} =100 \left(\frac{T_{\rm eff}}{5777K}\right)^{-32/9}\left(\frac{g}{g_{\odot}}\right)^{2/9}
\end{equation}

Similarly, the pulsation amplitude ratio in photometry
observations (ppm) and Doppler observations (m/s)\footnote{In practice, the wavelength of Doppler observations also affects the meansured RV amplitude. Here, these relations can be treated as the mean amplitude for the RVs meansured with the metal lines in the optical wavelength range.} is given by:
\begin{equation}
r_{puls,bolo}=\frac{a_{puls,bolo}(ppm)}{a_{puls}(m/s)} =20 \left(\frac{T_{\rm eff}}{5777K}\right)^{-\alpha}
\end{equation}

According to BC17, the theoretical value of $\alpha$ is 0.5, but in practice, a value closer to unity agrees better with observations. We adopt $\alpha=0.9$.
For observations in a certain passband, e.g. {\it Kepler}, we need to convert the bolometric variation to the variation in this filter. Following Burkart et al.\ (2012) and for a certain passband, we have
\begin{equation}
a_{puls,bolo}\beta(T_{\rm eff})= a_{puls,passband}
\end{equation}
where
\begin{equation}
\beta(T_{\rm eff}) \approx \frac{\int_{\lambda_1}^{\lambda_2}{(\partial B/\partial \ln T)d\lambda}}{4\int_{\lambda_1}^{\lambda_2}{Bd\lambda}}
\end{equation}

For example, for the Kepler passband, $\beta(T)=1.05$ when $T_{\rm eff}=6000$K and $B(\lambda)$ is the blackbody spectrum.

%$\beta(T)=0.81$ when $T_{\rm eff}=8500$K, 

Thus more generally, the amplitude ratio is given as:
\begin{equation}
r_{puls,passband}=\frac{a_{puls,passband}(ppm)}{a_{puls}(m/s)} =20\beta(T_{\rm eff}) \left(\frac{T_{\rm eff}}{5777K}\right)^{-\alpha}
\end{equation}

Thus, we can convert the photometric pulsation amplitude in {\it Kepler} passband $a_{puls,kepler}$ to the pulsation amplitude in Doppler observations: $a_{puls,RV}$:
\begin{equation}
\frac{a_{puls,kepler}}{a_{puls,RV}}=r_{puls,kepler}
\end{equation}

Since the height of the oscillation bump is measured in power: $H_{osc} \propto a_{puls}^2$, the height $H_{osc}$ in {\it Kepler} photometry $H_{osc, kepler}$ and RV $H_{osc,RV}$: is related by: 
\begin{equation}
\frac{H_{osc,kepler}}{H_{osc,RV}}=r_{puls,kepler}^2
\end{equation}

We assume that in photometry and Doppler observations the granulation frequencies ($b_1,b_2$) are the same. For the width of the oscillation bump ($\sigma_{env}$), we also assume that the width in RV is approximately the same as that in photometry: 
\begin{equation}
\sigma_{env,RV}\approx 1.0 \sigma_{env}.
\end{equation}
We refer to Houdek (1999) and the Appendix for a detailed discussion of this approximation (Fig.\ 14).

Thus, the RV counterpart of the photometric parameters $(b_1, b_2, a_1,a_2, H_{osc}, \sigma_{env})$ are $(b_1, b_2$, $a_{1,RV},a_{2,RV}$, $H_{osc,RV}$, $\sigma_{env,RV}$). 
In Table 1, we summarize all the scaling relations for the granulation and oscillations.

Starting from stellar parameters $\theta_{*}=(T_{\rm eff}, \log g)$, the simulated mean photometric (labeled as LC) and RV PDS are illustrated in the upper two panels of Figure 1. 
It can be seen that the oscillation bump is more dominant in RV observations than photometric observations. The `observed', noisy PDS (middle two panels) is generated from the mean PDS by drawing a sample from the underlying distribution of the Fourier power (see Sec. 3 for details). Finally, the PDS are converted to time series by performing the inverse-Fourier transform. The final products are the simulated RV (or photometric) time series as shown in the lower two panels of Fig. 1, with observing patterns (e.g., gaps and finite exposure times) added.

In the right panels of Figure 2, we show the simulated photometric PDS and RV PDS for three stars representing a main-sequence star (A), a sub-giant (B), and a giant (C). They have the following stellar parameters: ($T_{\rm eff}(K),\log g,\nu_{max}(\mu Hz)$)=(5777, 4.44, 3140),(5872, 3.95, 1020),(4986, 3.24, 213), respectively. The location of these three stars in the $T_{\rm eff}-\log g$ diagram is shown in the left panel. From A to C, the frequency at maximum power ($\nu_{max}$) decreases from $3140$ to $213\mu Hz$, and both the granulation and oscillation amplitude increase.

\onecolumngrid 

%46Lmi
\begin{figure}[htb!] 
\begin{center} 
{\includegraphics[height=12cm,angle=0]{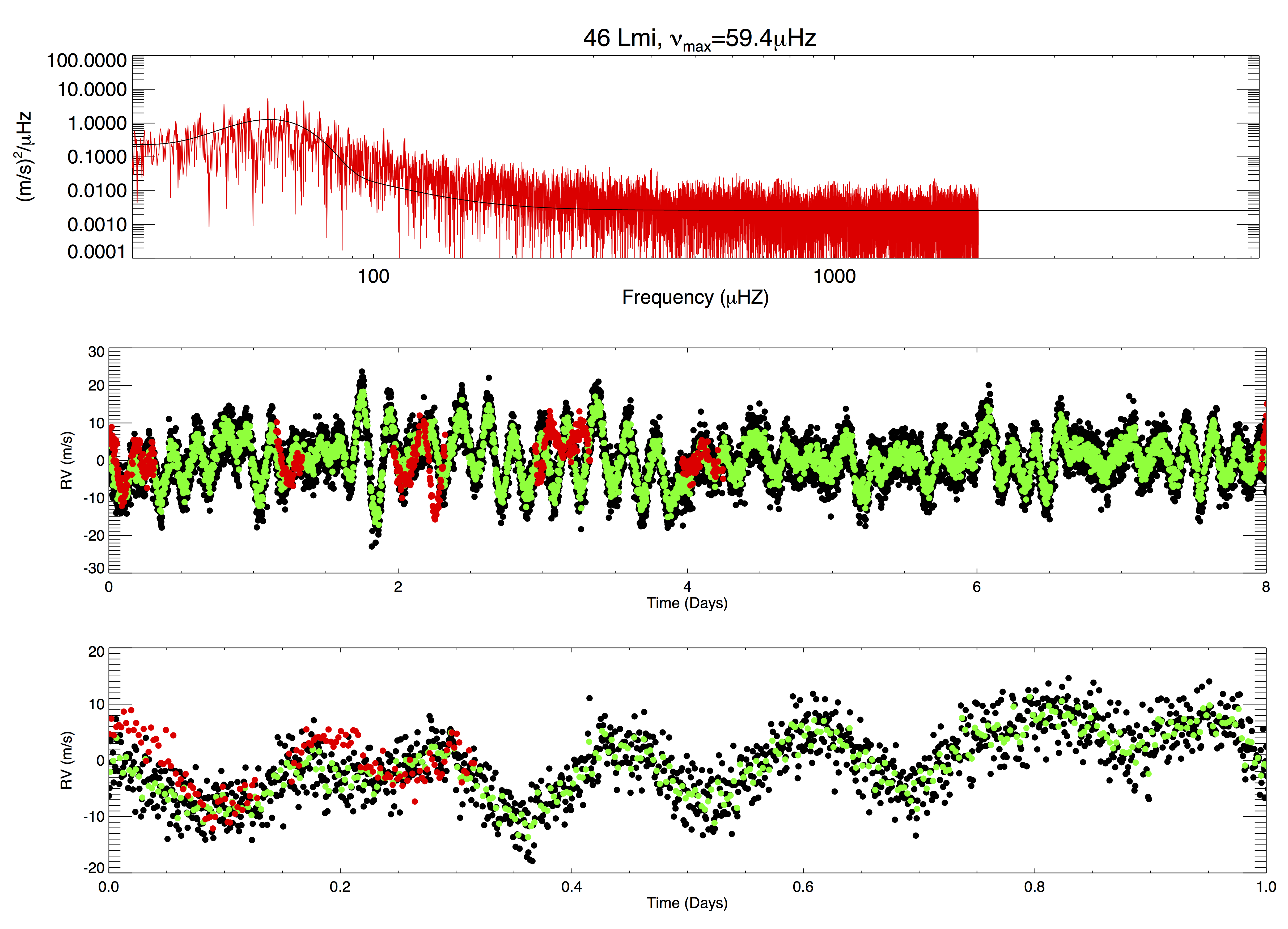}} 
\end{center} 
\caption{\textbf{Upper panel}: Our PDS model (black) and the observed RV PDS of 46 LMi from SONG (red); \textbf{Middle \& Lower panels}: Simualted RV time series with 1min cadence (black) and 5min cadence (green);  Observed RV time series with 5min cadence from SONG is overplotted in red. Note the similarity (i.e, amplitude, variation time scale) between the simulated (green) and observed (red) RV time series (they accidentally have simlar phases as well).}
\end{figure} 

%kapCrB
\begin{figure}[htb!]  
\begin{center} 
{\includegraphics[height=12cm,angle=0]{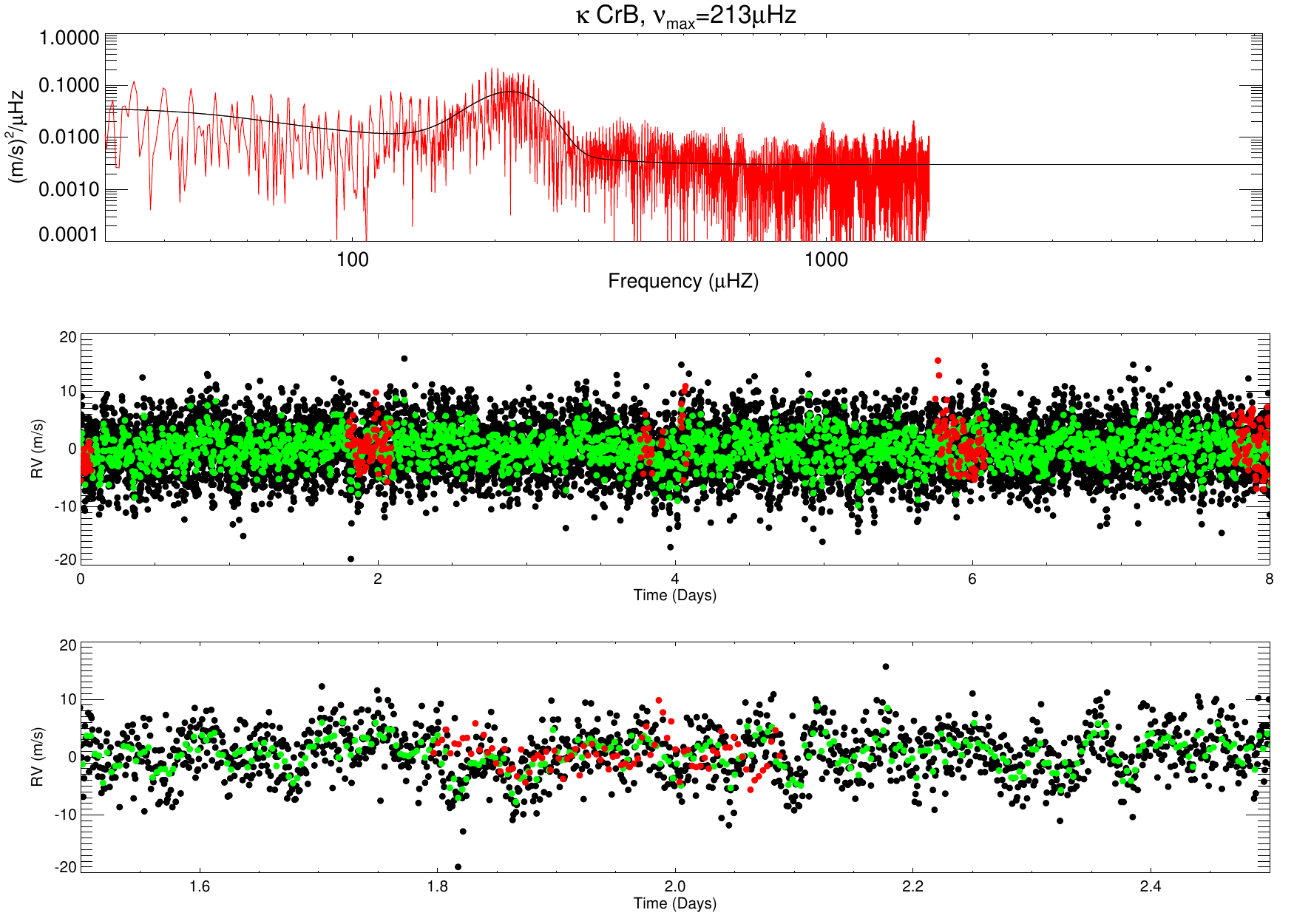}} 
\end{center} 
\caption{\textbf{Upper panel}: Our PDS model (black) and the observed RV PDS of $\kappa$ CrB from SONG (red); \textbf{Middle \& Lower panels}: Simulated RV time series with 1min cadence (black) and 5-min cadence (green);  Observed RV time series with 5-min cadence from SONG (red).  Note the similarity between the simulated (green) and observed (red) RV time series.}
\end{figure} 

%aCenA
%testsimuRV_chaplin.pro
\begin{figure}[htb!]  
\begin{center} 
{\includegraphics[height=12cm,angle=0]{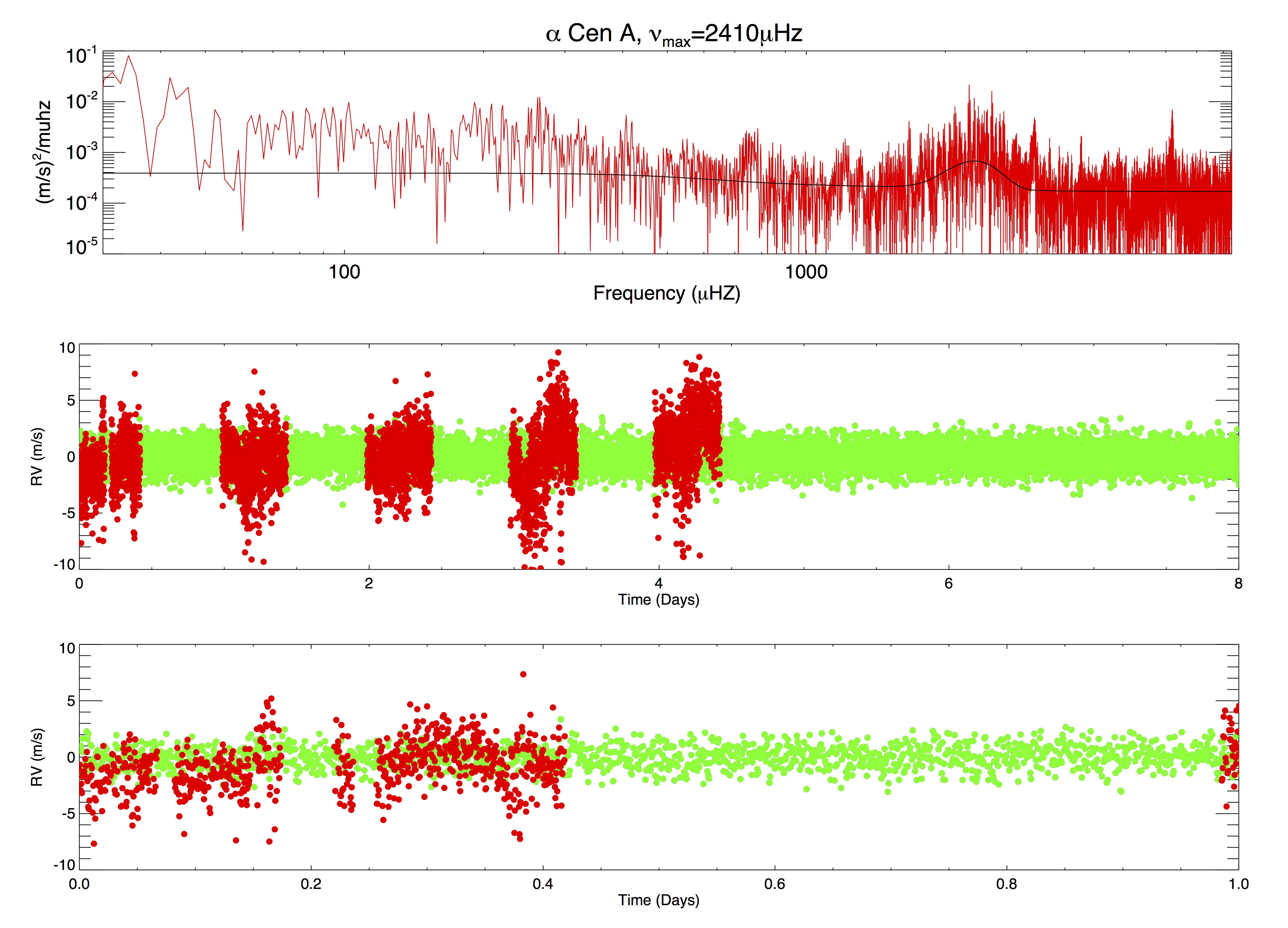}} 
\end{center} 
\caption{\textbf{Upper panel}: Our PDS model (black) and the observed RV PDS of $\alpha$ Cen A from HARPS (red); \textbf{Middle and Lower panels}: Simulated RV time series with 1min cadence (green);  Observed RV time series with 1min cadence from HARPS (red). The simulated (green) and observed (red) RV time series have similar amplitudes and variation time scales if the low frequency trend in the red RVs is removed (shown on the upper panel as frequencies larger than about 300 $\mu Hz$).}
\end{figure} 

\twocolumngrid 

\section{Simulating radial velocity time series with the inverse Fourier Transform}

The actual observed PDS can be modeled as the mean PDS multiplied by a random noise satisfying a $\chi^2$ distribution with 2 degree of freedom.
The observed power spectrum is denoted by $P_{\nu}$, and its probability density, $f(P_{\nu})$ is related to the mean PDS by (Handberg \& Campante 2011)
\begin{equation}
f(P(\nu;\Theta))=\frac{1}{\mathcal{P}(\nu;\Theta)}\exp{\left[-\frac{P(\nu)}{\mathcal{P}(\nu;\Theta)} \right]}
\end{equation}

To generate the observed PDS $P(\nu;\Theta)$ from the mean PDS $\mathcal{P}(\nu;\Theta)$, we draw a sample F from the uniform distribution defined on 0 to 1, and then solve for P from the cumulative distribution function of eq.(18): $F(P;\lambda)=1-e^{-\lambda P}$, where $\lambda$ is the parameter in the exponential distribution. Then we follow a similar procedure of Timmer \& Koenig (1995) for generating power-law noises in the AGN studies. For each frequency bin $\nu_j$ in the generated observed PDS $P(\nu;\Theta)$, we assign a complex variable with its amplitude equal to  $P(\nu_j;\Theta)$ and a random phase (a sample from the uniform distribution from $0$ to $2\pi$). Thus after an inverse Fourier transform to these complex variables, we obtain the corresponding time series in the time domain in flux (ppm) or RVs (m/s). The phase randomising does not change the RMS scatter of time series since the integrand in the Parseval's theorem only depends on the real part of the power spectrum.

The steps of simulating the mean photometric and radial velocity time series from the mean PDS are shown in the lower half of Figure 1.

In Figure 3 and Figure 4, We compare our simulated PDS and RV time series with those from observations. The observed RV data of two red giants 46 LMi, ($T_{\rm eff},\log g) = (4690K, 2.67)$ and $\kappa$ CrB, ($T_{\rm eff},\log g) = (4876K, 3.24)$ are obtained from the SONG (Stellar Observations Network Group)\footnote{https://phys.au.dk/song/}. In each figure, the upper panel shows the observed PDS in red and simulated mean PDS in black.
In the middle and lower panels, we compare the observed RV time series (red, with 5-min cadence) and simulated RV time series (black for 1-min cadence, green for 5min cadence) on two different timescales.
Comparing the red with the green RV time series, it can be seen that both the amplitude and time scale of the variation is well reproduced. 46 LMi shows variabilities mainly from the oscillation, as its RV time series presents periodicity of about 0.2 days, with amplitude about 10 m/s. For $\kappa$ CrB, both granulation and oscillation can be seen in the RV time series. The simulated RV time series (green) is very similar to the observed (red) time series.  Note that this is not a fit to the data, but rather a draw from a distribution. They accidentally have similar phases. It means we only expect the data (red) and our draw (green) to have similar statistical properties. Note also the gaps between observations in the red time series.

We show a similar comparison in Figure 5 for a main sequence star $\alpha$ Cen A. The upper plot shows that our PDS model underestimate the power at frequencies less than $\approx 200 \mu Hz$, this is why the simulated (green) RV time series seem to have lower amplitude than the observed (red) RVs. If the low frequency trend in each RV segment is removed, the two time series have comparable amplitude. There are various of factors that our simple simulation does not include, e.g., instrumental drift in RVs, super-granulation, rotational modulation and magnetic activity. It is thus not surprising that there are some differences between our model and real RV observations at low frequencies. This will require additional modeling which is beyond the scope of this paper.

\section{Root-mean-square RVs and Observation Strategies}    

\subsection{RMS RVs of stars at different evolutionary stages}

With the simulated RVs in hand, we examine the expected RMS RVs for stars with different evolutionary stages. We obtain three evolutionary tracks with masses of $0.8, 1.0, 1.2M_{\odot}$ from the MIST stellar models (Choi et al.\ 2016) . All models have solar metallicty and the ($T_{\rm eff},\log g)$ pairs are used to generate the simulated RVs as described in Sec. 2.2 and Sec. 3.

%Both granulation and oscillation comprise the `stellar jitter', which is the effect we need to mitigate in the exoplanet detection from precise RVs. Although stellar oscillations of solar-like stars are mainly from the power excess (oscillation bump) around $\nu_{max}$ with a typical width of $\frac{1}{2}\nu_{max}$ for main sequence and sub-giant stars (Chaplin \& Basu 2017), the frequency dependence of granulation (and super-granulation) spans several orders of magnitudes from minutes to days. The commonly referred 'RV jitter' is actually the integral of the power spectrum density from minimum frequency to the Nyquist frequency.

Our simulated RV time series have a cadence of 1-min and a time span of 17 days. To calculate the rms RV for a given integration time (20 min and 10min in Figure 6 and 7, respectively), we bin the original RVs and recalculate the standard deviation of the RVs.
The upper panels in Figure 6 and 7 show the RMS RVs for stars from ZAMS to the late-red giant stage, close to the tip of RGB. The lower panels focus on the main-sequence and sub-giant phases.
The RMS RVs on each track essentially increase as stars evolve from ZAMS to the red giants. 
%Also stars at same main-sequence phase but with higher masses tend to have higher RMS RVs. This is more obvious in the lower panel.

%[SURPRISING BECAUSE THE OSCILLATION AND GRANULATION AMPLITUDE SCALES ROUGHLY WITH L/M(?) AREN'T YOU JUST SEEING A DIFFERENE IN L/M...BECAUSE YOUR MORE MASSIVE STARS ARE MORE LUMINOUS FOR WHAT YOU CALL 'SAME EVOLUTION PHASE']. 

As mentioned above, we only consider the variability from stellar granulation and oscillation, our calculations here should be interpreted as lower-limit in the rms RVs. Other factors such as rotationally modulated stellar activity, long-term magnetic cycles, instrumental drift, etc are not considered. We also zero out the white noise term $W$ in the PDS since any RV time series possess an instrumental noise term that will vary from instrument to instrument. For practical use, one can add in quadrature an instrumental specific constant noise term to the reported RV rms in Figure 6 and 7.

\onecolumngrid 

%plotsigma_hr3.pro
\begin{figure}[ht]   
\begin{center} 
{\includegraphics[height=18cm,angle=90]{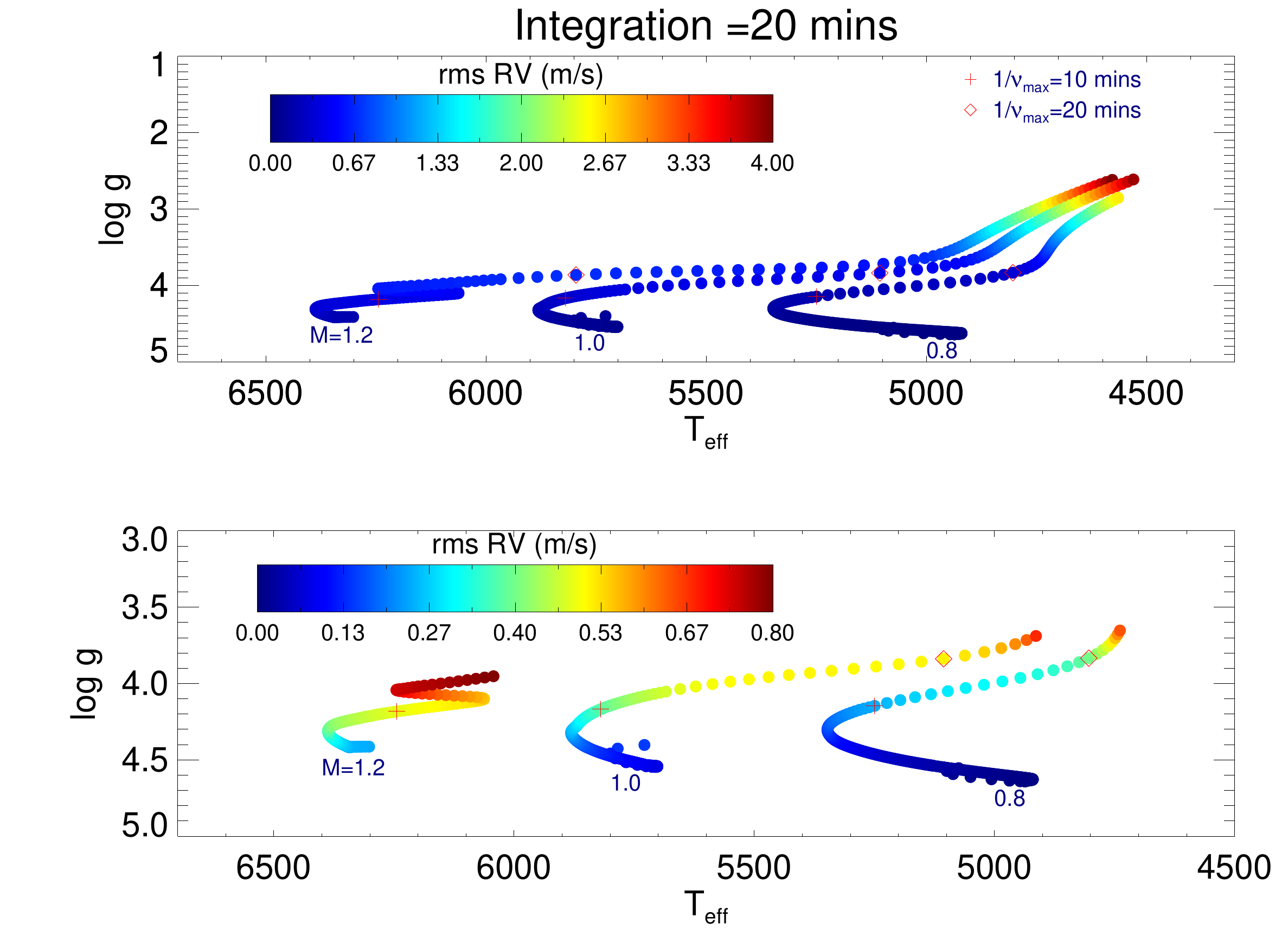}} 
\end{center} 
\caption{RV rms scatter for stellar models on three evolutionary tracks with $M=1.2, 1.0, 0.8M_{\odot}$. The integration time is 20 minutes. The lower panel focus on the less-evolved models. The RV rms scatter is color-coded, with increasing RV rms from blue to red. The red cross and diamond symbols on the evolutionary tracks indicate the locations where the stellar oscillation period ($1/\nu_{max}$) is equal to the integration time of 10 and 20 minutes, respectively. }
\end{figure} 

%plotsigma_hr3a.pro
\begin{figure}[ht]   
\begin{center} 
{\includegraphics[height=18cm,angle=90]{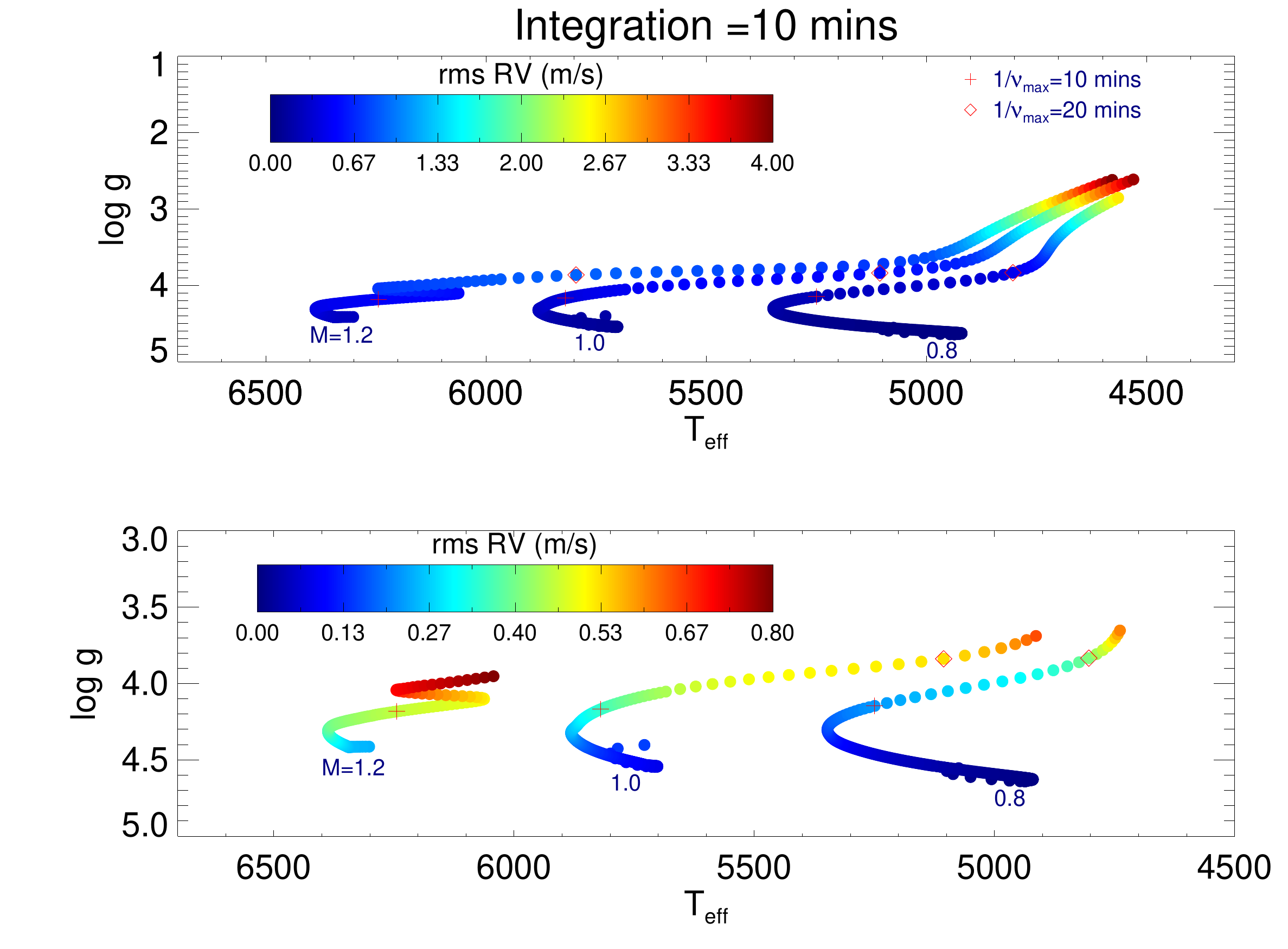}} 
\end{center} 
\caption{RV rms scatter for stellar models on three evolutionary tracks with $M=1.2, 1.0, 0.8M_{\odot}$. The RV rms scatter is color-coded, with increasing RV rms from blue to red. The integration time is 10 minutes. The lower panel focus on the less-evolved models.  }
\end{figure}

\twocolumngrid 

\subsection{The effect of observational strategies on the RV RMS}
 
In the exoplanet community, `RV jitter' is often used to refer to the RV excess after removing the Keplerian reflex motion of the star owing to the planet and accouting for reported RV measurement uncertainty. RV jitter values reported in the literature can be from different instruments with different instrumental noise properties. Even for the same instrument, different integration times and gaps in the time series will result in the RV jitter to be different. When comparing RV jitter for large samples of inhomogeneous measurements (Butler et al.\ 2017; Luhn et al.\ 2020), the aforementioned factors are rarely taken into account. Our method can account for these factors and can simulate realistic RV time series by introducing different integration times and gaps.

In Figure 8, we show the rms scatter of the RV time series as a function of the different integration time. The RV rms is a monotonically decreasing function of integration length. The characateristic decaying time is about the time scale of $1/\nu_{max}$, indicated by the red vertical line. The two blue vertical lines mark the characteristic timescale for the two granulation components in the PDS. When the integration length is longer than the granulation time scale, the RV rms begins to flatten out. Stars with different parameters would have a different RV rms constant at very long integration time. Note that there are NOT distinct local minima near $N/\nu_{max}$, as suggested by Chaplin et al.\ (2019) who considered only oscillations.

We show the effects of various survey strategies in Figure 9 using the stellar parameter of $\alpha$ Cen A. We simulate the RV time series with a sampling rate of 1min and then rebin the RVs to 7min to cancel out the oscillations (with dominate oscillation period$=1/(2410\mu Hz) \approx 7min$). We add daily gaps to the time series and only keep $N=1,2,3,4$, or $5$ RV data points in each night. These data points have a spacing of $dt=m\times 7$mins. $m=1$ corresponds to no gap between exposures. The cartoon in the right panel of Fig. 9 shows a RV time series with $N=3$. Finally, we calculate the RMS values of the nightly mean of these RV data points.
As mentioned in Section 3, each simulated RV time series has different phases depending on the different seed for the random number generator. To mitigate the effects of randomness, we average the resulting RMS of nightly-mean RVs over 200 different RV time series, where each simulation used a different seed. The left panel of Figure 9 shows the effect of the spacings between adjacent observations within one night on the RMS of nightly mean RVs. If we observe the star more than once per night, the nightly-mean RV rms values generally decrease as the spacings of data points increase. However, the RMS levels off once the spacing becomes larger than the characteristic timescale of granulation (when $m \ge 4$, i.e. when $dt$ is longer than $ \approx 28$ mins $\approx$ timescale of the lower-frequency granulation component of $\alpha$ Cen A). 

We  find  that  it  will  often  be  advantageous  to  introduce  gaps  in  observations  within a night comparable or longer than the granulation timescale.  If one were to neglect the effects of  granulation, then consecutive  observations would be expected to be more efficient, since once can avoid dead  time due to slewing  and acquiring the next target. However, for modern extremely precise radial velocity surveys, a larger spacing between observations  within a night may be a benefit larger enough to out-weight the cost of acquiring the target multiple times. For example,  the $N=  2,m= 3$ point is comparable to the $N= 3,m= 1$ point, and $N= 3,m= 3$ performs better than $N= 4,m= 1$. Once a telescope has left a target, there is not a significant increase in wall time required to increase the spacing (e.g., from $m= 2$ to $m= 5$).  Therefore, it may be advantageous to pick a set of $\approx 2$ to $5$ stars to cycle between two or more times within a time,  so as to suppress the effects of granulation.  Our model for generating realistic RV time series as a function of stellar properties, provides a valuable tool for observers seeking to optimize the efficiency of their observing strategy for an exoplanet survey given the properties of a particular target star (or set of stars).

\clearpage

\onecolumngrid 

\begin{figure}[ht!] 
\begin{center} 
{\includegraphics[height=18cm,angle=270]{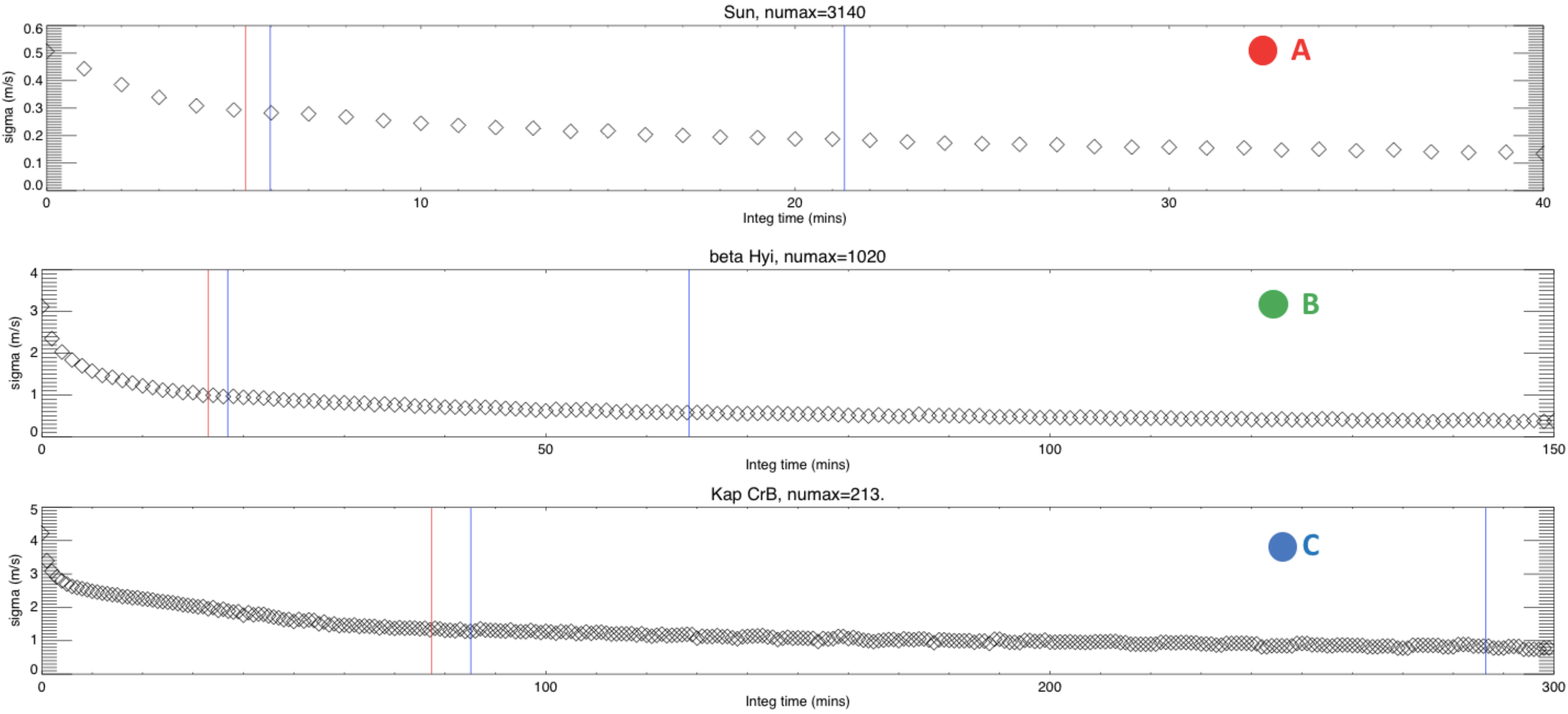}} 
\end{center} 
\caption{RV rms (m/s) as a function of integration length (minutes) for three stars and different evolutionary stages. A: main sequence, the Sun; B: sub-giant, $\beta$ Hyi, with $T_{\rm eff}=5872K, \log g=3.95$; C: giant, $\kappa$ CrB, with $T_{\rm eff}=4986K, \log g=3.24$. The red vertical line and two blue vertical lines label the characteristic time scale for stellar oscillation and two granulation components, respecively.}
\end{figure} 

\twocolumngrid 

\clearpage

\onecolumngrid 

\begin{figure}[ht!] 
\begin{center} 
{\includegraphics[height=11cm]{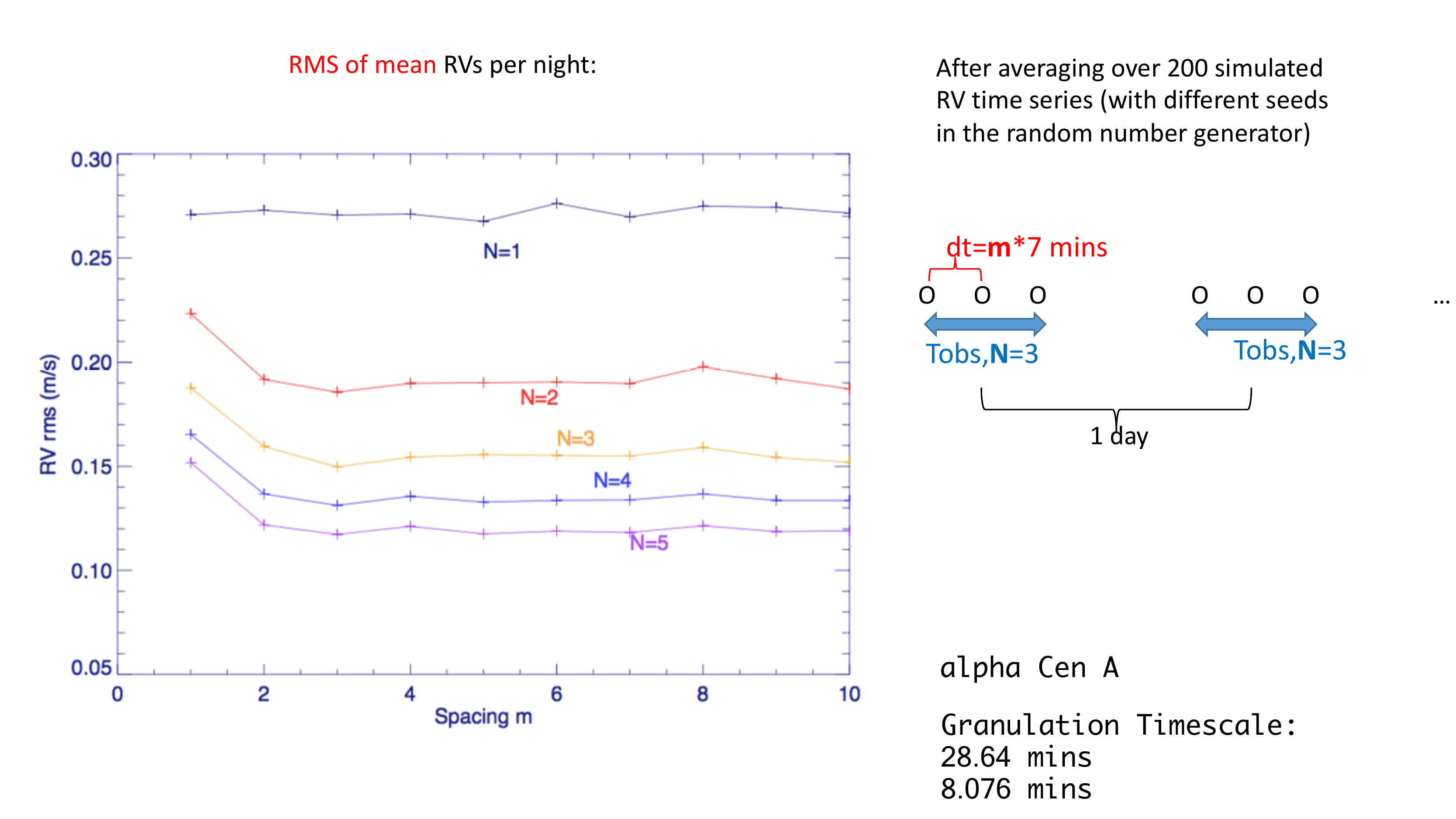}} 
\end{center} 
\caption{The observed RV time series as a function of the spacing between observations. Points along each curve are based on timeseries with N data points per night. Ajacent measurements are spaced by $dt= m \times 7$ mins, where $7$ mins $\approx 1/\nu_{max}$. The RMS of nightly-mean RVs are calculated by averaging over 200 simulated time series with different seeds in the random number generator.}
\end{figure} 

%What about adding a second panel for the Sun?  I ask because the granulation timescales for the sun are significantly longer, so the spacing needed to get a low RMS mightbe larger.] [EBF: Is there any difference in points for N=1?  If not, should that just be a horizontal line?

\twocolumngrid 

\section{Summary and Discussions}

Simply speaking, our simulation can be regarded as a black box, with either $\nu_{max}$ or the stellar parameter pair $(T_{\rm eff}, \log g)$ as inputs, and radial velocity time series (with any sampling rate or length) as an output. The simulation can be summarized as follows. Given $(T_{\rm eff}, \log g)$, we first calculate $\nu_{max}$ via Eq. (2). Then we obtain the granulation parameters ($b_1,b_2, a=a_1=a_2$) from equations 3, 4, 5 and oscillation parameters ($H_{osc}, \sigma_{env}$) from eq 8 and 9, respectively. Thus, the simulated photometric power spectrum density can be constructed via Eq. 1.
To obtain the corresponding RV PDS, we use Eq. 10(14) to convert the granulation and oscillation amplitudes in photometry to Doppler observations. This is a key step and it is based on the result suggested by BC17. Note that we still use Eq 3, 4 (granulation timescales are the same in RVs and in photometry) and eq. 9 (width of the oscillation bump) when converting to RV observations.

Our procedure incorporates the latest advancement in our understanding of stellar granulation and oscillation from {\it Kepler} photometric and asteroseismic stars in the Fourier space in K14. It differs from the approach by Luhn et al. (2020), which uses the empirical measurements of RV jitter from heterogeneous RV data sets (e.g., different integration time, gaps in the data).  Our simulated photometric time series can be added to the transit model to examine the effect of granulation/oscillation on the planet-size parameters (e.g. Sulis et al.\ 2020). Similarly, Keplerian orbits can be added to the simulated RVs to study the effect of `red noise' on the planet mass measurements (Gilbertson et al.\ 2020; Meunier \& Lagrange 2020).
Meunier et al.\ (2019) also used the power spectrum of granulation and oscillation to simulate RV time series, which follows the procedures in Dumusque et al.\ (2011, 2016).
Our methods differ from the above work in that we directly use the photometric calibrated granulation and oscillation amplitudes and explicitly show the scaling relations. We only consider the granulation and do not consider the super-granulation component used in Dumusque et al.\ (2011). This is because the so-called super-granulation component in RV PDS can be affected significantly by instrumental effect and long-term variabilities. We also find this component is not significant for sub-giants and giants, and its significance in the RV PDS of main sequence star requires further analysis. The work by Dumusque et al. only studied main sequence stars, whereas we also considered sub-giants and giant stars, whose granulation/oscillations are better-calibrated from the Kepler asteroseismic stars (K14).

Our method provides a mechanism for quantitatively prioritizing targets for upcoming radial velocity surveys based on the expected RV precision, taking into account the effects of granulation, pulsations and observing strategy.  Our method also provides a path for optimizing observing strategies for an extremely precise radial velocity survey and a given set of target stars.
Our method can be directly applied to stars with photometric data from {\it Kepler} or other missions such as {\it TESS}. The abundance of {\it TESS} photometric data enables the prediction and estimation of the corresponding granulation and oscillation signature in RVs for a large number of stars. When  high-quality photometry is not available or insufficient to measure $\nu_{max}$, our method can also be applied based on ($T_{\rm eff}, \log g$).  This can facilitate the target selection and the optimization of  observing strategy for a broad range of RV surveys. 

%This would facilitate the target selection and the observing strategy for designing RV surveys.

There are a number of caveats in this work:
1) It is possible to consider additional parameters for our model (e.g. [Fe/H]). Incorporating parameters other than $\nu_{max}$ could potentially make the model less reliant on the seismic scaling relations. 2) We approximate the stellar oscillations as a Gaussian-shaped power excess, ignoring the detailed structures (i.e., individual modes and their spacings). Including these details can change our results somewhat but not in a qualitative manner. 3) We only consider the granulation and oscillations, not stellar rotation and magnetic cycles. It is possible to parameterize stellar rotation periods as a function of age across the HR diagram, as is done in gyrochronology (Meibom et al.\  2015; van Saders \& Pinsonneault.\ 2013; van Saders et al.\ 2016, 2019). Stellar magnetic cycles, due to its long-term nature, is more difficult to quantify and parameterize. Subtle effects also include the magnetic effect on stellar oscillations, i.e., $\nu_{max}$ and p-mode amplitude and frequencies actually all vary with stellar cycles. Howe (2020) found that $\nu_{max}$ varies by 25 $\mu Hz$ through the solar cycle. P-mode power also slightly decreases as stellar activity strengthens (Santos et al.\ 2015). These variations arising from activity also apply to solar-like stars since the magnetic field is ubiquitous (Garc{\'\i}a et al.\ 2010). In addition, the oscillation power excess is also not strictly symmetric and deviates from a Gaussian shape (BC17). More detailed analysis is subject to further studies.
4) The measured granulation amplitude depends on the spectral lines used to measure the RVs. As lines are formed at different height in the photosphere, the resulting RV perturbations due to granulation and even oscillations may somewhat differ (BC17). In fact, Howe (2020) and Jimenez-Reyes (2007) showed that the line-formation region in the solar observations from VIRGO, GONG, BISON, GOLF blue, GOLF red have increasing atmospheric height, but actual observations do not follow this ordering, suggesting the way the observations are made also matters (see also Michel et al.\ 2009).

In addition, the two granulation components and one oscillation component can be parameterized in a way so their PDS expressions can have corresponding representation in the time domain, namely, the Gaussian process (GP) kernels. By using {\it Kepler} light curves and following Foreman-Mackey et al. (2017),  Pereira et al. (2019) showed that the PDS fitting in the Fourier domain for stellar granulation and oscillations is equivalent to the corresponding fitting in the time domain by using the celerite GP kernels. Barros et al.\ (2020) have already explored in this direction for photometric time series by adding GP kernels for stellar rotation as well. Kunovac Hod{\v{z}}i{\'c} et al.\ (2020) used celerite-GP kernels to account for the granulation and oscillation perturbation to RV time series, which is crucial for revealing the underlying Rossiter-Mclaughlin (RM) signal buries in stellar variabilities. We defer to a future paper to predict the GP kernel parameters that can be used for modeling the photometric and Doppler observations from stellar parameters.

%osc/granulation RV amplitudes depend on the line formation depth, Chaplin \& Basu (2017),

%TODO: find the relation between celerite kernel parameters and K14 scaling relation parameters

\appendix

\section{RV Power Spectrum Fitting }  

%\textbf{plots of psd fitting}

We have assumed that the PDS of RV time series can be modeled by two Lorentzian functions for granulation and one Gaussian function for oscillation, similar to photometric PDS. Here, we test this assumption by fitting the observed RV-PDS of stars with extensive quasi-continuous RV observations from HARPS and SONG and examining if they satisfy the similar scaling relations. 

Generally speaking, the Fourier spectrum of solar-like oscillating stars consists of a background component (granulation, super-granulation, stellar activity), an oscillation bump (acoustic modes) and a flat component (white noise). 
In asteroseismology, it is a common practice to fit the background of the power spectrum density by using some functions (Lorentzian and super-Lorentzian). This background component is subtracted before the peak-bagging procedure (frequency extraction). Early studies of the power spectrum of solar RV time series in Harvey (1985) used three Lorentzian functions with index of 2: $P(\nu)=4\sigma^2 \tau/(1+(2\pi\nu\tau)^2)$. Lefebvre et al.\ (2008) used two components of the same functions to fit PDS background of the SOHO/GOLF RV time series.

Although these are some physical reasons behind these functional forms (e.g., sudden increase or exponentially increasing signal with exponential decaying, BC17), it is more empirical than physical.  For example, Mathur et al.\ (2011) presented the modeling of PDS background by index of 2, 4 and a combination of both. Michel et al.\ (2009) used two Lorentzian components with the power index of $\nu$ as a free parameter for the photometric power spectrum in different passbands.
K14 focused on the {\it Kepler} photometric observations. They performed Bayesian model comparison and argued that two super-Lorentzian functions with index=4 are the best model describing the PDS background of {\it Kepler} light curves. The functional form is: 
\begin{equation}
P(\nu)=\sum_{i=1,2}\frac{\xi_{i}a_{i}^2/b_i}{(1+(\nu/b_i)^4)}
\end{equation}.

Later, Kallinger et al.\ (2019) analyzed the granulation signals in the red giant light curves observed by the BRITE satellite. They found that one or two super-Lorentzians are needed to fit the background, depending on the quality of the data.

Similarly, we can work on the PDS of the RV time series. These have been extensive ground-based spectroscopic campaign dedicated to finding solar-like oscillations. Most of the work focus on detecting oscillations and little attention is paid to the extraction of the granulation signal in the power spectrum. 
Dumusque et al.\ (2011) is an exception. This work fitted the PDS background by using three components of the form $A_i/(1+(B\nu_{i}))^{C_i}$, representing the super-granulation, meso-granulation\footnote{The reality of meso-granulation is still debated.}, and granulation. The index $C_i$ is regarded as a free parameter, and the fitting result of $C$ ranges from 2.6 to 19.8! 

To directly compare the Doppler and photometric granulation amplitude, we use the super-Lorentzian functions in K14 to fit the power spectrum of our RV time series.
We choose stars with extensive RV observations: $\alpha$ Cen A, $\alpha$ Cen B, $\beta$ Hyi, $\mu$ Ara from the HARPS (Dumusqure et al.\ 2011), and $\gamma$ Ceph, $\kappa$ CrB, 6 Lyn, 46 LMi from the SONG spectrograph (Stello et al.\ 2017). We also use SOHO/GOLF observations of the Sun.

We fit the PDS background with three models, corresponding to one, two, and three super-Lorentzian components. We then determine the best model following the model comparison method in de Assis Peralta (2018) (see also Appourchaux et al.\ 1998; Karoff 2012).
Briefly, the power spectrum density satisfies a $\chi^2$ distribution and let us compare two models A and B. Model A has two super-Lorentzian components ($p$ parameters) and Model B  has three components ($p+q$ parameters). we calculate the logarithmic likelihood ratio $\ln\Lambda=\mathcal{L}(\lambda_{p+q})-\mathcal{L}(\lambda_{p})$, and $\ln\Lambda$ follows a  $\chi^2$ distribution with two degree of freedom (DOF). We use the metric $-2 \ln \Lambda$ to determine if the additional component is significant.
We set $\Delta L (a\%)$ as the a\% quantile of the $\chi^2$ distribution with $DOF=2$.  If  $-2 \ln \Lambda \leq \Delta L$, the additional component in model B is not significant, and we adopt model A; if  $-2 \ln \Lambda \geq +\Delta L$, model B is significant.

%Downloads/psd_rv/
\begin{figure}[htb!] 
\begin{center} 
{\includegraphics[height=12cm]{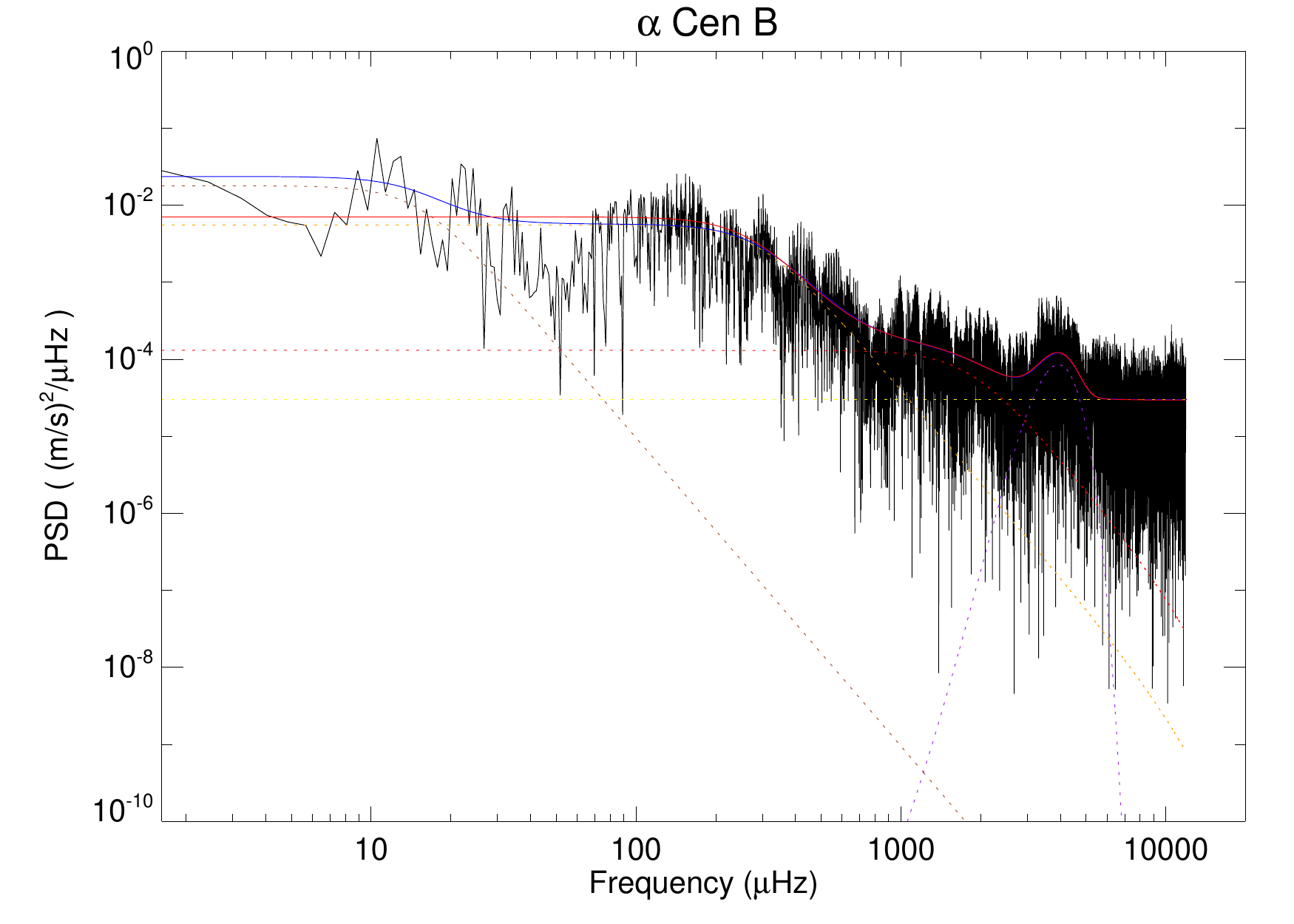}} 
\end{center} 
\caption{The PDS fitting of the RVs of $\alpha$ Cen B. The three-granulation-component model (Blue) and two-granulation=component model (Red) can both fit the observed data well, and a model comparison is needed. The four dotted curves show the PDS of each of the component signals.}
\end{figure}

%/stars_for_zhao/plotpsd_GamCeph1.pro
% Fig 2, fitting results for the rest 9 stars
\begin{figure}[htb!]  
\begin{center} 
{\includegraphics[height=12cm]{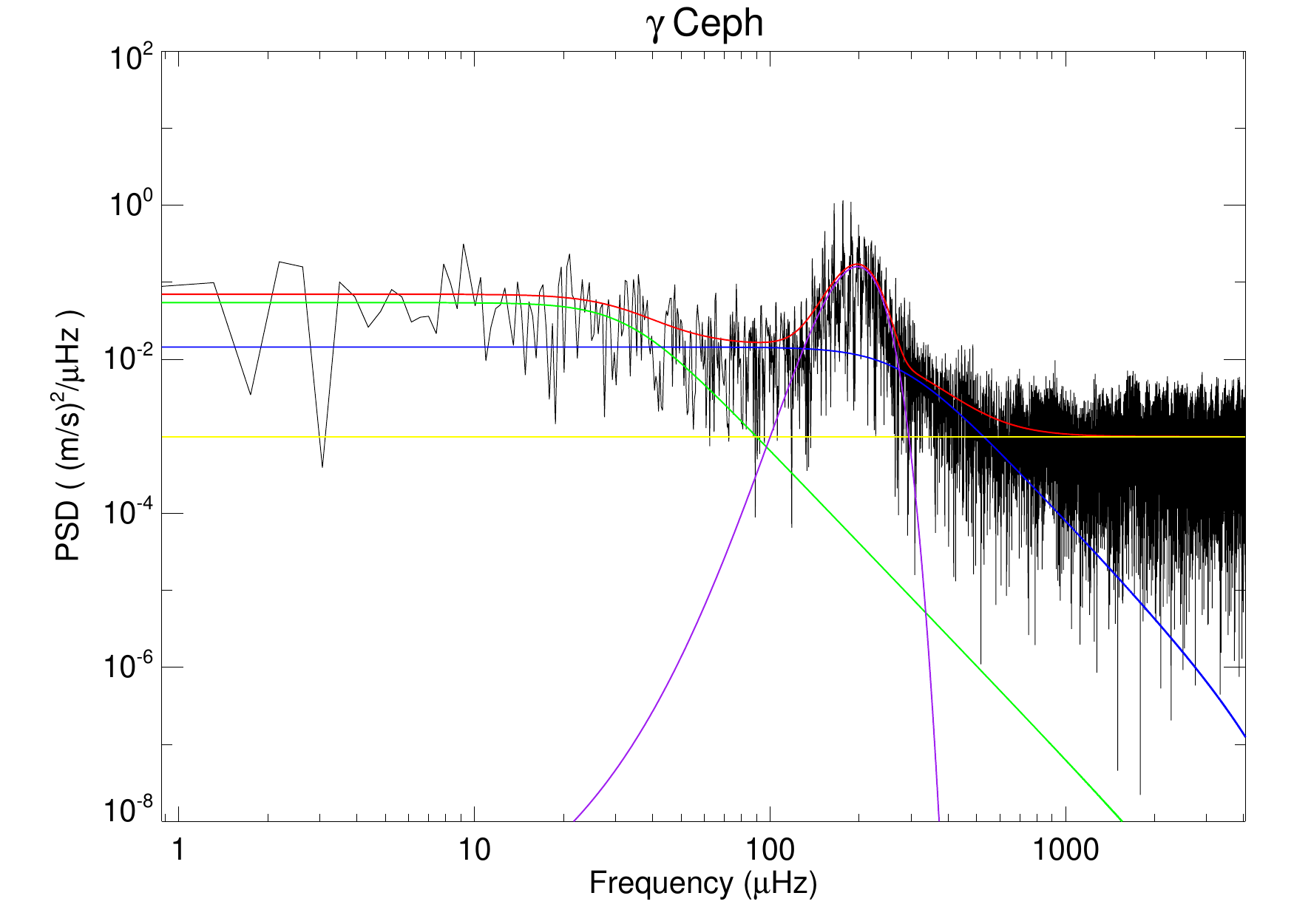}} 
\end{center} 
\caption{The PDS fit for $\gamma$ Ceph is shown as the red line. The two super-lorentzian functions are indicated by the green and blue lines. The yellow line marks the constant Gaussian noise.}
\end{figure}

%/Users/fffeynman/Downloads/stars_for_zhao/ratio.pro
\begin{figure}[htb!]  
\begin{center} 
{\includegraphics[height=12cm]{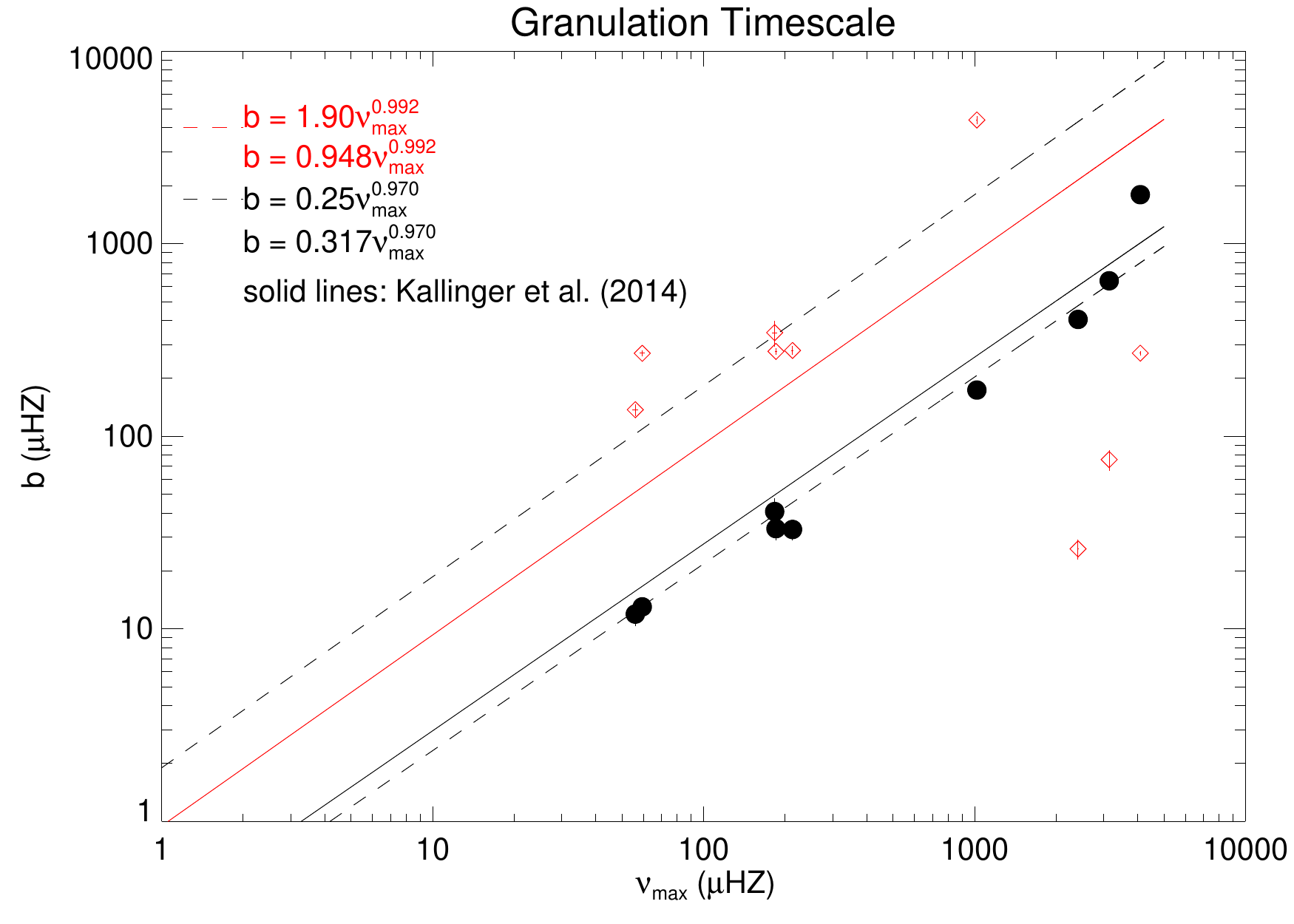}} 
\end{center} 
\caption{Two-component granulation model fit to the observed RV PDS. We find that, for sub-giants and red giants, the two components have similar timescale to the K14 result; For main sequence stars, we find the low-frequency component in phtometric PDS in K14 can be recovered in the RV data. The observed RV PDS favor the existence of a separete low-frequency component model, which is probably due to another mechanism for contributing to the low-frequency variability in addition to granulation. }
\end{figure} 

%Fig3, rv_gran vs numax
\begin{figure}[htb!]  
\begin{center} 
% {\includegraphics[angle=90,height=12cm]{flcsample.eps}} 
{\includegraphics[height=12cm]{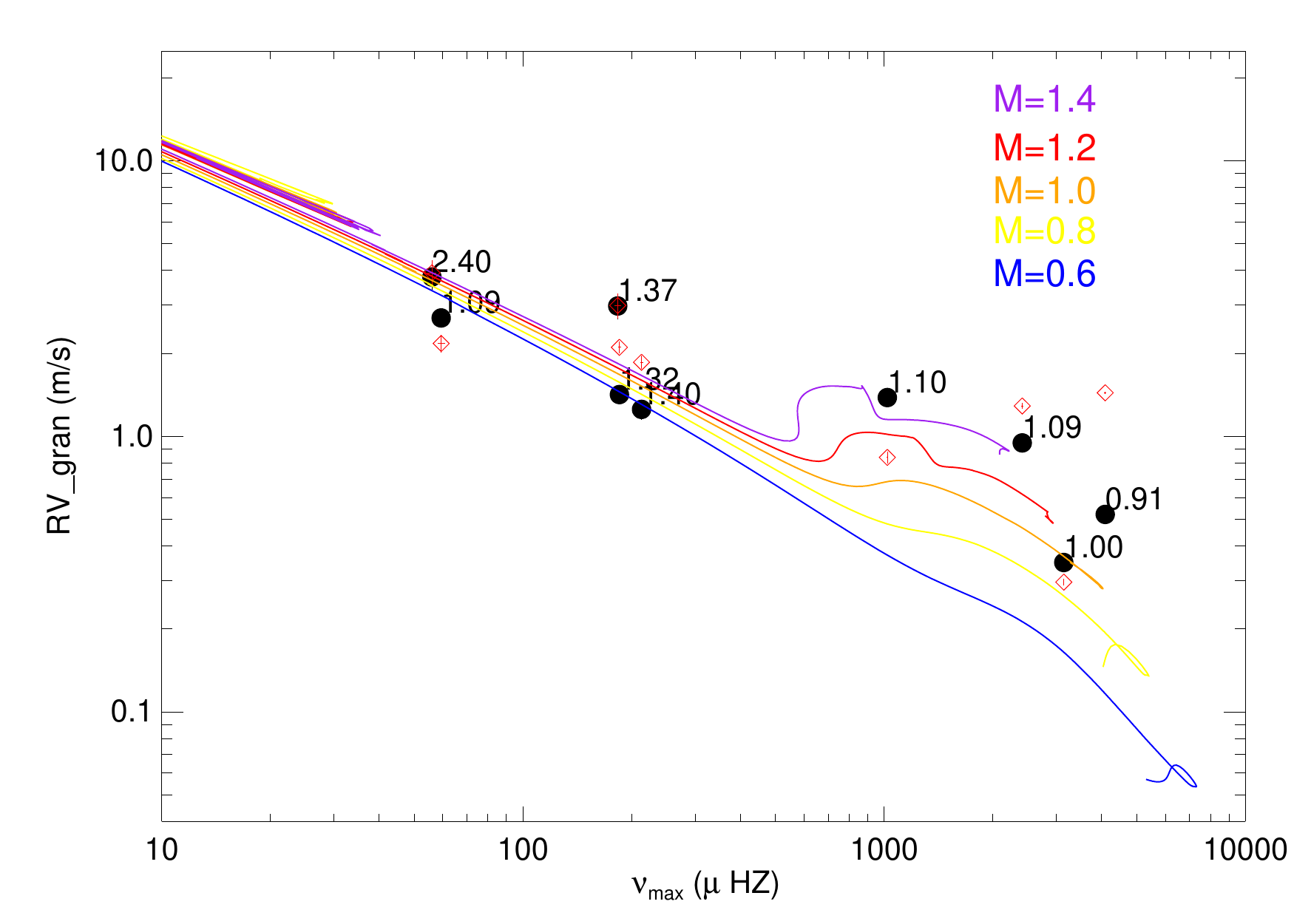}} 
\end{center} 
\caption{Measured RV amplitudes of the two granulation components $RV_{gran}=a_1, a_2$(dots and diamonds, respectively) as a function of $\nu_{max}$. The stellar masses in solar units are labeled. Lines in color correspond to the theoretical granulation amplitude in RVs for a series of stellar structure models calculated with the MESA evolution code. The tracks span the mass range from $0.6$ to $1.4$ with solar metallicity $Z=0.02$. Stars evolve from the lower right to the upper left.}
\end{figure}

\begin{figure}[htb!]  
\begin{center} 
{\includegraphics[height=11cm]{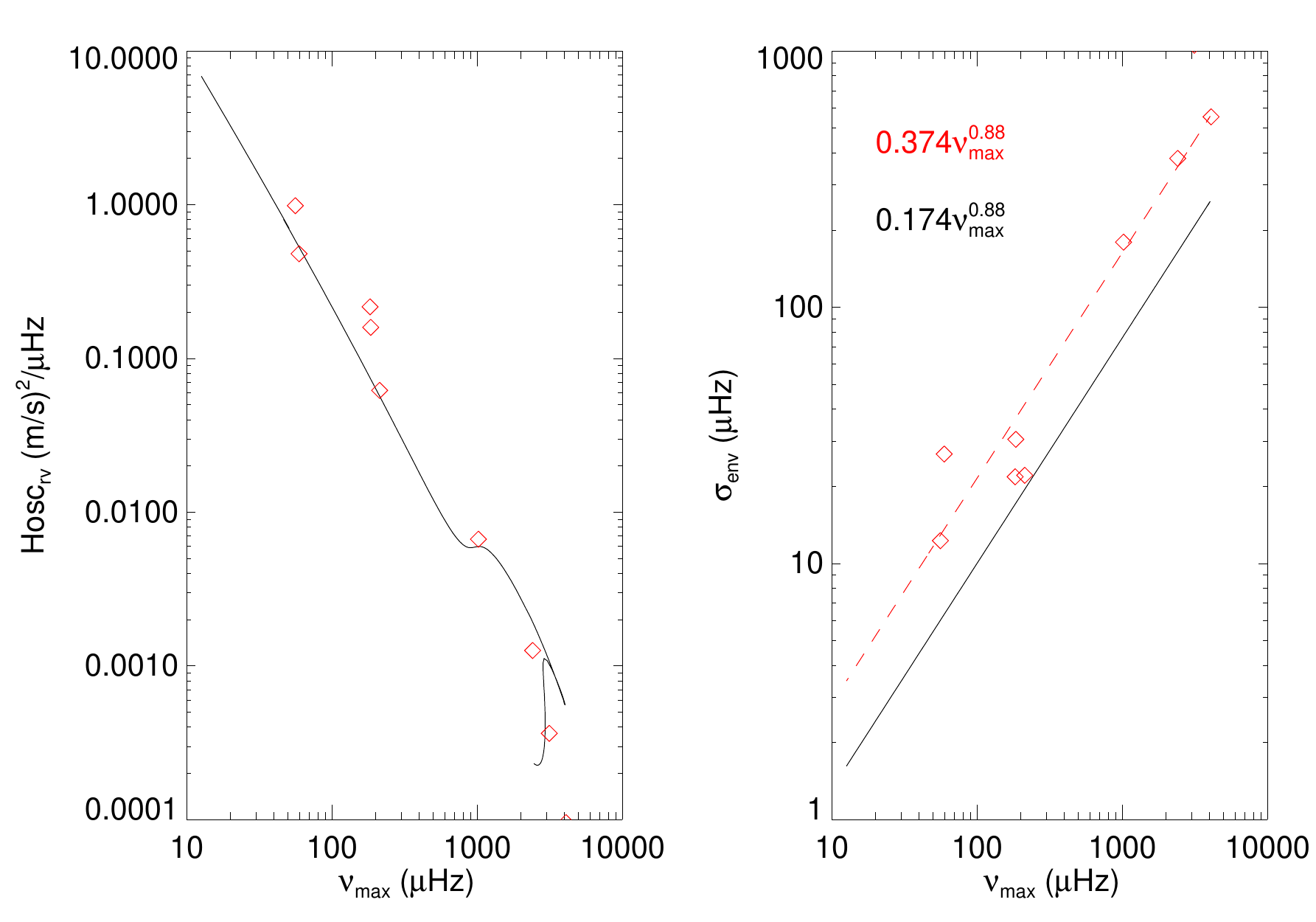}} 
\end{center} 
\caption{\textbf{Left}: The height of the Gaussian-shaped oscillation bump ($H_{osc,rv}$) in the RV PDS as a function of $\nu_{max}$. \textbf{Right}: The width of the Gaussian-shaped oscillation bump in the RV PDS as a function of $\nu_{max}$. Diamonds indicate the measurements from the observed RV PDS and the two lines indicate theoretical predictions from two different scaling laws.}
\end{figure} 

An example is presented in Figure 10. We show the best-fitting PDS background model with two components (model A) and three components (model B) for $\alpha$ Cen B. We find that  $-2 \ln \Lambda= -56$, which is much smaller than $-\Delta L (99\%)=-9.21$. Thus the third super-Lorentzian component is not needed and we adopt the two-component model. We find the same result for most stars and thus adopt the two-granulation-component model. An example fit to the PDS of $\gamma$ Ceph is shown in Figure 11.  The advantage of adopting the two-component model is that the results for RV PDS can be directly compared to the result for photometric PDS in K14.  Table 2 summarises our results.

We show the two granulation time scale parameters $b_1, b_2$ in RVs for our stars in Figure 12. We find that the lower frequency component from the PDS fitting, essentially follows the expected scaling relation from K14: $b_1=0.317\nu_{max}^{0.97}$. The knee frequency of the high-frequency granulation component also fellow the K14 relation but only for the giants and sub-giants in this sample with $\nu_{max} <  1500\mu Hz$. For the main sequence stars, we obtain essentially the same low-frequency component as in K14 (the $b_1$ component), but an additional lower-frequency component is required. This is in line with Figure 5, in which we find the K14 two-component granulation model underestimates the variability at low frequencies.
It suggests that a low-frequency variability is dominant which is not accounted for in the K14 model. Factors contributing to this low-frequency variability include the instrumental effect, rotational modulation, and magnetic activity.

The corresponding amplitude parameters $a=a_1,a_2$ from fitting the RV PDS are plotted in Figure 13 (labeled as $RV_{gran}$ on the y-axis). We over-plot the theoretical granulation amplitude calculated from the scaling relations used in Section 2 (also Table 1). We find the RV granulation amplitude is well predicted by the scaling relations. Evolutionary tracks of different masses are from MIST modes with solar metallicity. 

In the left panel of Figure 14, we compare the measured oscillation height $H_{osc}$ from the RV PDS fitting with the theoretical expectations (eq.\ 8, 16) (Table 1). We find good agreement with the K14 scaling relations. The right panel of Figure 14 contains the measured width ($\sigma_{env}$) of the oscillation bump.  We find our measured width is about a factor of two larger than the photometric scaling relations used in Section 2. More specifically, the bump width from {\it Kepler} photometry scales as  $0.174\nu_{max}^{0.88}$ (eq.\ 9, black line in Fig. 14). However, in Doppler observations, the bump width is about two times wider, satisfying a scaling of $0.374\nu_{max}^{0.88}$ (red dashed line). This difference arises from the fact that the RV time series contains gaps while the {\it Kepler} photometric time series is almost continuous. The aliases introduced in the window function partly contribute to the measured bump width in the RV PDS.  We thus still assume the same bump width in the RV PDS when generating RV time series in Section 2.

%%%%%%%%%%%%%%%%%%%%%%%%%%%%%%%%%%%%%%%%%%%%%%%%%%%%%%%%%%%%%%% 
 
\acknowledgments

This work was supported by funding from the Center for Exoplanets and Habitable Worlds.  The Center for Exoplanets and Habitable Worlds is supported by the Pennsylvania State University, the Eberly College of Science, and the Pennsylvania Space Grant Consortium.

%{\it Facilities:} \facility{Kepler, Mayall} 
 
%%%%%%%%%%%%%%%%%%%%%%%%%%%%%%%%%%%%%%%%%%%%%%%%%%%%%%%%%%%%%%% 
% References 

%%%%%%%%%%%%%%%%%%%%%%%%%%%%%%%%%%%%%%%%%%%%%%%%%%%%%%%%%%%%%%% 

\clearpage

%%%%%%%%%%%%%%%%%%%%%%%%%%%%%%%%%%%%%%%%%%%%%%%%%%%%%%%%%%%%%%% 

% Figures  old

% testsimuRV_plot_betaHyi.pro
% Fig 2, fitting results for the rest 9 stars
%\begin{figure} 
%\begin{center} 
%{\includegraphics[height=10cm]{psdflow.pdf}} 
%\end{center} 
%\caption{The flow chart of simulating photometric and Doppler time series from the stellar power spectrum density (PSD).}
%\end{figure} 

%Downloads/ arvind.pro
%\begin{figure} 
%\begin{center} 
%{\includegraphics[height=12cm,angle=0]{hr_psd_newcut.pdf}} 
%\end{center} 
%\caption{Simualted power spectrum density of three representative stars (Black: Photometric PSD in ${(ppm)}^2/ \mu Hz$; Red: Doppler PSD in ${(m/s)}^2/ \mu Hz$). A: main sequence, the Sun  ($T_{\rm eff}=5777K, \log g=4.44, \nu_{max}=3140\mu Hz$); B: sub-giant, $\beta$ Hyi ($T_{\rm eff}=5872K, \log g=3.95,\nu_{max}=1020 \mu Hz$; C: giant, $\kappa$ CrB ($T_{\rm eff}=4986K, \log g=3.24,\nu_{max}=213 \mu Hz$). It is apparent that the Gaussian-shaped oscillation bump is more dominant in the Doppler PSD.}
%\end{figure} 

%testsimuRV_chaplin.pro
%\begin{figure} 
%\begin{center} 
%{\includegraphics[height=12cm,angle=0]{Sun_simu.eps}} 
%\end{center} 
%\caption{\textbf{Upper panel}: Our PSD model (black) and the observed solar RV PSD from GOLF (red); \textbf{Middle and Lower panel}: Simualted RV time series of the Sun with 1min cadence; observed GOLF RVs of the Sun with 1min cadence. }
%\end{figure} 

\clearpage

\begin{deluxetable}{cccccccc}
%c is center aligned, l is left aligned, r is right aligned
\tabletypesize{\small} 
%\rotate 
\tablewidth{0pc} 
\tablenum{1} 
\tablecaption{Scaling relations for the granulation and oscillation\label{tab1}} 
\tablehead{ 
\colhead{-}        &      
\colhead{Parameters (units)}          & 
\colhead{Photometry (Kepler)}&
\colhead{Doppler}        &                \\  
\colhead{}           & 
\colhead{}           & 
\colhead{(Kallinger 2014)}                 &
\colhead{(this work)} &        
} 
\startdata		
\textbf{Granulation}: &    &       &       \\
 Amplitude&    $a_1$, $a_2$ (ppm)  &  $a=3382(9)\nu_{max}^{-0.609(2)} (Eq. 5) $     & Eq. 10      \\
 Frequency&        $b_1$ ($\mu Hz$)&     $b_1=0.317(2)\nu_{max}^{0.970(2)}$   & $b_1=0.25\nu_{max}^{0.97}$              \\
Frequency&       $b_2$ ($\mu Hz$)&     $b_2=0.948(3)\nu_{max}^{0.992(2)}$   &    $b_2=1.90\nu_{max}^{0.992}$           \\
\hline
\textbf{Oscillation}: &    &       &       \\
Power excess width&        $\sigma_{env}$ ($\mu Hz$)&     $\sigma_{env}=0.174\nu_{max}^{0.88}$    & $\sigma_{env,RV}=0.374\nu_{max}^{0.88}$          \\
 Power excess height&       $H_{osc}$ ($10^3$ ppm$^2/\mu Hz$)&     $H_{osc}=11800 \nu_{max}^{-2.3}$   &     Eq. 16        \\
\enddata 
\end{deluxetable}

%$a=3382(9)\nu_{max}^{-0.609(2)} $
%$H_{osc}=11120(1900)\nu_{max}^{-2.21(5)}$  

%use /Downloads/stars_for_zhao/psdrv_result.txt
 \begin{deluxetable}{cccccccccccc}
%c is center aligned, l is left aligned, r is right aligned
\tabletypesize{\footnotesize} 
\rotate 
\tablewidth{0pc} 
\tablenum{2} 
\tablecaption{RV PDS fitting parameters\label{tab1}} 
\tablehead{ 
\colhead{Name}          & 
\colhead{W}&
\colhead{$a_1$}        & 
\colhead{$b_1$}        & 
\colhead{$a_2$}    &
\colhead{$b_2$}        &                    
\colhead{$H_{osc}$}           & 
\colhead{$\nu_{max}$}           & 
\colhead{$\sigma_{env}$}           &  \\
\colhead{}                 &
\colhead{((m/s)$^2\mu$Hz$^{-1}$)} & 
\colhead{(m/s)}& 
\colhead{($\mu$HZ)} & 
\colhead{(m/s)}& 
\colhead{($\mu$HZ)}& 
\colhead{((m/s)$^2\mu$Hz$^{-1}$)}&  
\colhead{($\mu$HZ)} & 
\colhead{($\mu$HZ)} &       
} 
\startdata		
Sun &  5.226e-06 $\pm$ 3.239e-08 &    0.349   $\pm$0.008  &    642.5  $\pm$    19.8  &   0.30 $\pm$   0.02&      75.8  $\pm$    9.4  & 0.000365
 $\pm$ 0.000006  &    3780.2   $\pm$  11.3   &   1054.1      $\pm$ 7.7 \\
$\alpha$ Cen B&   2.95586e-05$\pm$  3.89041e-07 &      0.521$\pm$    0.011 &       1799.1$\pm$      63.1 &       1.44$\pm$    0.03 &       270.58$\pm$
      7.34 &   9.589e-05$\pm$  5.3e-06 &       3938.2$\pm$      24.8 &       554.0$\pm$      37.2\\
$\alpha$ Cen A&   0.00023$\pm$  2.6e-06 &      0.947$\pm$    0.031 &       404.93$\pm$      21.25 &       1.29$\pm$     0.13 &       26.1$\pm$
      3.1 &    0.001263$\pm$  6.36e-05 &       2418.8$\pm$      14.1 &       381.3$\pm$      10.6\\
$\beta$ Hyi&   9.59101e-05$\pm$  3.81970e-06 &       1.383$\pm$    0.042 &       174.0$\pm$      6.5 &      0.8386$\pm$    0.0264 &       4394.7$\pm$
      225.2 &    0.00669$\pm$  0.00034 &       1018.6$\pm$      5.3 &       179.6$\pm$      3.9\\
$\gamma$ Ceph&   0.000994813$\pm$  1.17809e-05 &       1.418$\pm$     0.106 &       33.18$\pm$      4.39 &       2.105$\pm$    0.096 &       276.2$\pm$
      11.1 &      0.160$\pm$    0.015 &       197.5$\pm$      1.7 &       30.53$\pm$      1.57\\
$\kappa$ CrB&    0.003039$\pm$  6.879e-05 &       1.253$\pm$     0.103 &       32.86$\pm$      4.13 &       1.858$\pm$    0.076 &       279.4$\pm$
      14.8 &     0.0621$\pm$   0.0086 &       211.8$\pm$      2.1 &       22.11$\pm$      1.78\\
6 Lyn&     0.01374$\pm$   0.00106 &       2.973$\pm$     0.317 &       40.704$\pm$      6.932 &       2.99$\pm$     0.35 &       345.67$\pm$
      53.96 &      0.217$\pm$    0.051 &       176.9$\pm$      3.9 &       21.82$\pm$      3.07\\
46 LMi&    0.00262$\pm$  2.6775e-05 &       2.69$\pm$     0.16 &       13.025$\pm$     0.919 &       2.173$\pm$    0.049 &       270.14$\pm$
      10.52 &      0.4808$\pm$    0.0223 &       57.12$\pm$      1.42 &       26.75$\pm$      1.11\\
$\epsilon$ Tau&    0.0029995$\pm$  5.614e-05 &       3.79$\pm$     0.42 &       11.94$\pm$      1.63 &       3.924$\pm$     0.279 &       137.55$\pm$
      9.75 &      0.99$\pm$     0.19 &       56.40$\pm$      1.54 &       12.29$\pm$      1.43\\
\enddata 
\end{deluxetable}


\clearpage 
 
\begin{thebibliography}{} 

\bibitem[Appourchaux et al.(1998)]{1998A&AS..132..107A} Appourchaux, T., Gizon, L., \& Rabello-Soares, M.-C.\ 1998, \aaps, 132, 107

\bibitem[Barros et al.(2020)]{2020A&A...634A..75B} Barros, S.~C.~C., Demangeon, O., D{\'\i}az, R.~F., et al.\ 2020, \aap, 634, A75

\bibitem[Basu \& Chaplin(2017)]{2017asda.book.....B} Basu, S., \& Chaplin, W.~J.\ 2017, Asteroseismic Data Analysis: Foundations and Techniques

\bibitem[Blancato et al.(2020)]{2020arXiv200509682B} Blancato, K., Ness, M., Huber, D., et al.\ 2020, arXiv:2005.09682

\bibitem[Bugnet et al.(2018)]{2018A&A...620A..38B} Bugnet, L., Garc{\'\i}a, R.~A., Davies, G.~R., et al.\ 2018, \aap, 620, A38

\bibitem[Butler et al.(2017)]{2017AJ....153..208B} Butler, R.~P., Vogt, S.~S., Laughlin, G., et al.\ 2017, \aj, 153, 208

\bibitem[Burkart et al.(2012)]{2012MNRAS.421..983B} Burkart, J., Quataert, E., Arras, P., et al.\ 2012, \mnras, 421, 983

\bibitem[Chaplin et al.(2019)]{2019AJ....157..163C} Chaplin, W.~J., Cegla, H.~M., Watson, C.~A., et al.\ 2019, \aj, 157, 163

\bibitem[Choi et al.(2016)]{2016ApJ...823..102C} Choi, J., Dotter, A., Conroy, C., et al.\ 2016, \apj, 823, 102

\bibitem[Corsaro et al.(2017)]{2017A&A...605A...3C} Corsaro, E., Mathur, S., Garc{\'\i}a, R.~A., et al.\ 2017, \aap, 605, A3. doi:10.1051/0004-6361/201731094

\bibitem[de Assis Peralta et al.(2018)]{2018AN....339..134D} de Assis Peralta, R., Samadi, R., \& Michel, E.\ 2018, Astronomische Nachrichten, 339, 134

\bibitem[Delisle et al.(2020)]{2020A&A...635A..83D} Delisle, J.-B., Hara, N., \& S{\'e}gransan, D.\ 2020, \aap, 635, A83. doi:10.1051/0004-6361/201936905

\bibitem[Dumusque et al.(2011)]{2011A&A...525A.140D} Dumusque, X., Udry, S., Lovis, C., et al.\ 2011, \aap, 525, A140

\bibitem[Dumusque(2016)]{2016A&A...593A...5D} Dumusque, X.\ 2016, \aap, 593, A5

\bibitem[Foreman-Mackey et al.(2017)]{2017AJ....154..220F} Foreman-Mackey, D., Agol, E., Ambikasaran, S., et al.\ 2017, \aj, 154, 220

\bibitem[Garc{\'\i}a et al.(2010)]{2010Sci...329.1032G} Garc{\'\i}a, R.~A., Mathur, S., Salabert, D., et al.\ 2010, Science, 329, 1032. doi:10.1126/science.1191064

\bibitem[Gilbertson et al.(2020)]{2020arXiv200901085G} Gilbertson, C., Ford, E.~B., Jones, D.~E., et al.\ 2020, arXiv:2009.01085

\bibitem[Handberg \& Campante(2011)]{2011A&A...527A..56H} Handberg, R., \& Campante, T.~L.\ 2011, \aap, 527, A56

\bibitem[Harvey(1985)]{1985ESASP.235..199H} Harvey, J.\ 1985, Future Missions in Solar, Heliospheric \& Space Plasma Physics, 235, 199

\bibitem[Houdek et al.(1999)]{1999A&A...351..582H} Houdek, G., Balmforth, N.~J., Christensen-Dalsgaard, J., et al.\ 1999, \aap, 351, 582

\bibitem[Howe et al.(2020)]{2020MNRAS.493L..49H} Howe, R., Chaplin, W.~J., Basu, S., et al.\ 2020, \mnras, 493, L49

\bibitem[Jim{\'e}nez-Reyes et al.(2007)]{2007ApJ...654.1135J} Jim{\'e}nez-Reyes, S.~J., Chaplin, W.~J., Elsworth, Y., et al.\ 2007, \apj, 654, 1135


\bibitem[Kallinger et al.(2014)]{2014A&A...570A..41K} Kallinger, T., De Ridder, J., Hekker, S., et al.\ 2014, \aap, 570, A41 

\bibitem[Kallinger et al.(2016)]{2016SciA....2E0654K} Kallinger, T., Hekker, S., Garcia, R.~A., et al.\ 2016, Science Advances, 2, 1500654

\bibitem[Kallinger et al.(2019)]{2019A&A...624A..35K} Kallinger, T., Beck, P.~G., Hekker, S., et al.\ 2019, \aap, 624, A35

\bibitem[Karoff(2012)]{2012MNRAS.421.3170K} Karoff, C.\ 2012, \mnras, 421, 3170

\bibitem[Kjeldsen \& Bedding(1995)]{1995A&A...293...87K} Kjeldsen, H. \& Bedding, T.~R.\ 1995, \aap, 293, 87

\bibitem[Kjeldsen \& Bedding(2011)]{2011A&A...529L...8K} Kjeldsen, H. \& Bedding, T.~R.\ 2011, \aap, 529, L8. doi:10.1051/0004-6361/201116789

\bibitem[Kunovac Hod{\v{z}}i{\'c} et al.(2020)]{2020arXiv200711564K} Kunovac Hod{\v{z}}i{\'c}, V., Triaud, A.~H.~M.~J., Cegla, H.~M., et al.\ 2020, arXiv:2007.11564

\bibitem[Lefebvre et al.(2008)]{2008A&A...490.1143L} Lefebvre, S., Garc{\'\i}a, R.~A., Jim{\'e}nez-Reyes, S.~J., et al.\ 2008, \aap, 490, 1143

\bibitem[Luhn et al.(2020)]{2020AJ....159..235L} Luhn, J.~K., Wright, J.~T., Howard, A.~W., et al.\ 2020, \aj, 159, 235

\bibitem[Mathur et al.(2011)]{2011ApJ...741..119M} Mathur, S., Hekker, S., Trampedach, R., et al.\ 2011, \apj, 741, 119

\bibitem[Meibom et al.(2015)]{2015Natur.517..589M} Meibom, S., Barnes, S.~A., Platais, I., et al.\ 2015, \nat, 517, 589. doi:10.1038/nature14118

\bibitem[Meunier et al.(2019)]{2019A&A...627A..56M} Meunier, N., Lagrange, A.-M., Boulet, T., et al.\ 2019, \aap, 627, A56

\bibitem[Meunier \& Lagrange(2020)]{2020arXiv200811952M} Meunier, N. \& Lagrange, A.~M.\ 2020, arXiv:2008.11952

\bibitem[Michel et al.(2009)]{2009A&A...495..979M} Michel, E., Samadi, R., Baudin, F., et al.\ 2009, \aap, 495, 979

\bibitem[Pereira et al.(2019)]{2019MNRAS.489.5764P} Pereira, F., Campante, T.~L., Cunha, M.~S., et al.\ 2019, \mnras, 489, 5764

\bibitem[Santos et al.(2016)]{2016MNRAS.461..224S} Santos, A.~R.~G., Cunha, M.~S., Avelino, P.~P., et al.\ 2016, \mnras, 461, 224

\bibitem[Stello et al.(2017)]{2017MNRAS.472.4110S} Stello, D., Huber, D., Grundahl, F., et al.\ 2017, \mnras, 472, 4110

\bibitem[Sulis et al.(2020)]{2020A&A...636A..70S} Sulis, S., Lendl, M., Hofmeister, S., et al.\ 2020, \aap, 636, A70

\bibitem[Tayar et al.(2019)]{2019ApJ...883..195T} Tayar, J., Stassun, K.~G., \& Corsaro, E.\ 2019, \apj, 883, 195. doi:10.3847/1538-4357/ab3db1

\bibitem[Timmer \& Koenig(1995)]{1995A&A...300..707T} Timmer, J., \& Koenig, M.\ 1995, \aap, 300, 707

\bibitem[van Saders \& Pinsonneault(2013)]{2013ApJ...776...67V} van Saders, J.~L. \& Pinsonneault, M.~H.\ 2013, \apj, 776, 67. doi:10.1088/0004-637X/776/2/67

\bibitem[van Saders et al.(2016)]{2016Natur.529..181V} van Saders, J.~L., Ceillier, T., Metcalfe, T.~S., et al.\ 2016, \nat, 529, 181. doi:10.1038/nature16168

\bibitem[van Saders et al.(2019)]{2019ApJ...872..128V} van Saders, J.~L., Pinsonneault, M.~H., \& Barbieri, M.\ 2019, \apj, 872, 128. doi:10.3847/1538-4357/aafafe

\bibitem[Yu et al.(2018a)]{2018MNRAS.480L..48Y} Yu, J., Huber, D., Bedding, T.~R., et al.\ 2018, \mnras, 480, L48. doi:10.1093/mnrasl/sly123

\bibitem[Yu et al.(2018b)]{2018ApJS..236...42Y} Yu, J., Huber, D., Bedding, T.~R., et al.\ 2018, \apjs, 236, 42. doi:10.3847/1538-4365/aaaf74
\end{thebibliography}
\end{document}